\documentclass[12pt]{iopart}
\pdfoutput=1

\usepackage{setstack}
\usepackage{iopams}
\eqnobysec

\usepackage{eurosym}
\usepackage{amsthm}
\usepackage{booktabs}
\usepackage{bbm}
\usepackage{hyperref}
\usepackage{graphicx}
\usepackage{ytableau}
\usepackage{cite}

\newtheorem*{prop}{Proposition}
\newtheorem*{theo}{Theorem}
\def\={\ =\ }

\newcommand{\ii}{\mathrm{i}}
\newcommand{\dd}{\mathrm{d}}

\begin{document}

\title[Phase transition in complex-time Loschmidt echo]{Phase transition in complex-time Loschmidt echo of short and long range spin chain}
\author{Leonardo Santilli}
\address{Grupo de F\'{i}sica Matem\'{a}tica, Departamento de Matem\'{a}tica, Faculdade de Ci\^{e}ncias, Universidade de Lisboa, Campo Grande, Edif\'{\i}cio C6, 1749-016 Lisboa, Portugal.}
\ead{\mailto{lsantilli@fc.ul.pt}}

\author{Miguel Tierz$^{1,2}$}
\address{$^1$ Departamento de Matem\'{a}tica, ISCTE - Instituto Universit\'{a}rio de Lisboa, Avenida das For\c{c}as Armadas, 1649-026 Lisboa, Portugal.}
\address{$^2$ Grupo de F\'{i}sica Matem\'{a}tica, Departamento de Matem\'{a}tica, Faculdade de Ci\^{e}ncias, Universidade de Lisboa, Campo Grande, Edif\'{\i}cio C6, 1749-016 Lisboa, Portugal.}
\ead{\mailto{mtpaz@iscte-iul.pt}, \mailto{tierz@fc.ul.pt}}

\begin{abstract}
We explain and exploit the random matrix formulation of the Loschmidt echo
for the XX spin chain, valid for multiple domain wall initial states and also for
a XX spin chain generalized with additional interactions to more neighbours. For models with interactions decaying as $e^{-\alpha \left\vert l-j\right\vert }/\left\vert l-j\right\vert ^{p+1}$, with $p$ integer or natural number and $\alpha \geq 0$, we show that there are third order phase transitions in a scaling limit of the complex-time Loschmidt echo amplitudes. For the long-range version of the chain, we use an exact result for Toeplitz determinants with a pure Fisher-Hartwig singularity, to obtain exactly the Loschmidt echo for complex times and discuss the associated Stokes phenomena. We also study the case of a finite chain for one of the generalized XX models.
\end{abstract}

\maketitle

\section{Introduction}

A central theme in modern physics is the study of out of equilibrium
properties, especially in systems of many-body physics. Remarkable progress
has been made, especially in recent years, in part due to the
realization that many out of equilibrium properties actually admit
descriptions analogous to the much better established case of
equilibrium physics. This analogy holds in some settings closer than initially expected. Chief among this family of
developments is the study of dynamical quantum phase transitions (DQPT) \cite%
{HPK,Heyl:2018abn}, which qualitatively can be thought of as phase transitions in time.  

A simple but yet very relevant protocol to study DQPT's is that of
quantum quenches, in which a system is prepared in some well-defined state
and it is left to evolve, unitarily, with some Hamiltonian $\hat{H}$. This
provides a fruitful setting for the exploration of many-body physics out of
equilibrium. This procedure can be encapsulated in the analysis of the
Loschmidt echo, which is a global quantity, and is defined as the squared
absolute value of the overlap between evolved and initial quantum states. It
is a simple and natural quantity to evaluate after a quantum quench. Given
the initial state, the Loschmidt amplitude is%
\begin{equation}
\mathcal{G}(t)=\langle \psi _{0}|\psi _{0}(t)\rangle =\langle \psi
_{0}|e^{-\mathrm{i} \hat{H} t}|\psi _{0}\rangle \,,  \label{Amp}
\end{equation}%
and, correspondingly, the Loschmidt echo is%
\begin{equation}
\mathcal{L}(t)=\big|\mathcal{G}(t)\big|^{2}\,.  \label{L}
\end{equation}%

A relevant and complete approach to the understanding of the real time dynamics encoded
in \eref{Amp} involves considering the time extended into the complex plane, $\ii t \mapsto w= \beta + \ii t \in \mathbb{C} $, 
and to study the resulting partition function \cite{HPK,Heyl:2018abn}%
\begin{equation}
\mathcal{Z} (w)=\langle \psi _{0}|e^{-w \hat{H}}|\psi _{0}\rangle ,  \label{theZw}
\end{equation}%
which leads to many possibilities, involving analytical continuations and
the consideration of Fisher and/or Yang-Lee zeroes \cite{HPK,Heyl:2018abn}. 
In general, when considering \eref{theZw}, the problem of validity of
analytical continuations will emerge, since a possible starting point
is to consider real values of $w$, which corresponds to imaginary time
evolution, where analytical results are sometimes more feasible and then
analytic continuation is used to obtain the real-time evolution of the
Loschmidt echo. The validity of this procedure is then a relevant problem
in itself. For example, in \cite{Piroli:2017mmz}, this procedure was shown, for
the case of the gapped regime of the XXZ spin chain model, to provide the
correct result only up to a finite time $t^{\ast }$. In general, a Wick
rotation will miss Stokes phenomena \cite{Stephan2017} and indeed, we will
give an example of a model where the analytical evaluation of \eref{theZw}
holds for $w\in \mathbb{C}$ and where this consideration of Stokes phenomena is made completely
explicit.

We will study Loschmidt amplitudes of spin chain models, a subject with already many analytical results \cite{Piroli:2017mmz,Stephan2017,PE,Piroli:2018amn,Piroli:2017sei,BP,Fzeroes,Echofermion,qreturn,Largedev}.
Our approach here will be based on the fact that the Loschmidt amplitude of XX
spin chains, defined below, admits a random matrix description, following
the discussion in \cite{DM}, although this specific result was originally
worked out in \cite{Bog,Bog2,B,BC}, but not in the context of Loschmidt
echo. These works then are in a slightly different setting and no discussion
of Loschmidt echo is made there. Thus, this random matrix formulation is not
discussed in the now large literature on these quantum amplitudes, with the
exception of particular cases involving very specific states with one domain
wall \cite{qreturn}. The mathematically equivalent formulation in terms of
Toeplitz (for periodic boundary conditions) and Toeplitz+Hankel determinants
(for open boundary conditions) is known \cite{Echofermion}, as it is the
equivalent descriptions in terms of Fredholm determinants \cite{Largedev}.

The determinantal representation describes a very specific family of the
amplitudes studied in \cite{Bog,Bog2,B,BC} and \cite{DM}. 
Determinants, as explained in \cite{DM} and as we shall see below, describe
the situation where the initial state is of the type%
\begin{equation}
|\psi _{0}\rangle =\vert \underset{N}{\underbrace{\downarrow ,\downarrow ,...,\downarrow }},\uparrow ,...\rangle ,
\end{equation}%
which we shall refer to as single-domain wall configuration (even
though, for example for periodic boundary conditions there are two domain
walls), whereas the generalization of Toeplitz minors \cite{Bog,Bog2,B,BC} and \cite{DM}
describes initial states with an arbitrarily complex pattern of spin
flips, such as, say 
\begin{equation}
|\psi _{0}\rangle =\vert \downarrow ,\uparrow ,\uparrow ,\downarrow ,\uparrow ,\downarrow ,\uparrow ,...\rangle ,
\end{equation}%
named multi-domain wall configuration. Notice that we write down
these examples denoting an infinite spin chain, but at the end of work we
will describe, as was done in \cite{DM}, the finite chain case.

These multi-domain wall configurations are described by minors of the same
Toeplitz or Toeplitz+Hankel matrices, as explained in \cite{DM}. From the
random matrix theory point of view, which provides an integral
representation of such minors \cite{Minor}, this generic multi-domain wall configuration leads to random
matrix ensembles with insertions of Schur polynomials, a family of symmetric
polynomials which are also characters of the unitary group \cite{RM} (see
section \ref{app:multidomain}). Interestingly, the same identity that establishes the
equivalence between Toeplitz determinants and unitary matrix model
ensembles, namely Andreief identity \cite{RM,qreturn}, also leads to the
equivalence between Toeplitz minors and unitary matrix models with Schur
polynomial insertions \cite{Minor}. In addition, the valuable Szeg\H{o}
theorem for the behavior of large determinants is extended to the case of
minors as well \cite{Minor}.

It is worth mentioning that the determinantal and random matrix formulation
given for quantum return probabilities, using free fermions, of \cite{qreturn}
coincides with that for the XX chain, obtained by N. M. Bogoliubov and
collaborators, including the study of asymptotic limits (leading to a
Gaussian random matrix ensemble) and different boundary conditions, studied in both \cite%
{qreturn} and \cite{B}. The latter analysis includes in addition the above
mentioned description of multiple spin flips in the initial states. 

It is also noteworthy that, while the random matrix description is in
principle rather specific to the XX model, it turns out that the case where
additional interactions are added to the XX chain can be studied in
the same fashion \cite{PGT}. An analogous statement is briefly made also in 
\cite[Appendix D]{Echofermion}, for their equivalent study of the free
fermionic chain, with single-domain wall initial configuration.

To have a random matrix description of a physical quantity is useful in many
ways. One of them is precisely to establish the existence of a phase
transition, normally in the context of a double-scaling limit, where the
size of the matrix model is taken to infinity at the same time as the
physical parameter, keeping their product (or ratio) constant. This was the
case of the celebrated Gross-Witten-Wadia phase transition \cite{GW,Wadia},
which has later on reemerged in many other areas and is still subject of
attention. The double scaling limit of such model is sophisticated, involving
Tracy-Widom law and solutions of Painlev\'{e} equations, and became central in
mathematics through the seminal work \cite{BDJ} and the universality of the
model and its scaling \cite{BDS}.

In the present work, we generalize the XX chain in different ways, and prove that the resulting systems undergo phase transitions exploiting the matrix model description. We first consider a spin chain with only nearest neighbour interaction, but now we allow the strength of the interaction to be different on the left and on the right. After that, we study the case in which each spin interacts with all the others, with interaction between two spins placed $n$ sites apart on the chain decaying either exponentially or as $1/n$ (or more generally as $1/n^{1+p}$).

Before proceeding, it is important to emphasize the differences between these matrix model phase transitions
(which are ubiquitous and central in establishing phase transitions in gauge
theories, see \cite{GW,Wadia,MinwallaWadia,AmadoSundborg} for example) and the DQPT's. While
the DQPTs occur when the time-evolved state $|\psi (t)\rangle $ becomes
orthogonal to the initial state vector $|\psi (0)\rangle $, which can be
studied by analyzing Fisher zeroes of \eref{theZw}, the matrix model phase
transitions imply that a rescaled version of $\mathcal{Z} (w)$ will have discontinuous
derivatives of some order (typically, third) in such a way
that this complex-time rescaled Loschmidt echo will posses at least two different phases, according to the rescaled value of the parameter.\par
\medskip
The material is organized as follows. In the next section we review the matrix model description of transition amplitudes on infinitely long spin chains, with possibly multi-domain wall initial configuration. 
Then, in section \ref{app:multidomain} we provide exact results for the multi-domain wall, as well as formulae for the asymptotic behaviour. 
Section \ref{app:saddleptsmethod} is devoted to calculations in the asymptotic regime, using random matrix theory, and contains the proofs of the phase transitions discussed in sections \ref{sec:GWasym} and \ref{sec:XXextended}. 
Section \ref{sec:GWasym} contains the analysis of the modified XX chain, in which the interaction is different on the left and on the right. We call it the XX chain with asymmetric hopping. We study the representation as Toeplitz determinants, and the corresponding asymptotics, both for imaginary and real time. 
In section \ref{sec:XXextended} we consider more general spin interactions, both short and long range. Through the matrix model representation, we prove the presence of phase transitions in the regime when the size of the matrix becomes large. The generalized interactions considered are: exponentially decaying (subsection \ref{sec:interactionBaik}), decaying as $1/n$ (subsection \ref{sec:interactionFH}), being $n$ the distance between the two spins, and eventually the case of more general decay (subsection \ref{sec:GenIntP}). In subsection \ref{sec:interactionFH}, we also provide exact formulae for the evaluation of the amplitudes for long-range interaction.
Finally, in section \ref{sec:finite} we discuss the case of a finite chain and explain how the arguments presented in previous sections are modified.

\section{Spin chains and matrix models}

The $S=1/2$ Heisenberg XX spin chain is a very well-known integrable
magnetic chain \cite{LSM}. This infinite chain
(which we consider with periodic boundary conditions) is characterized by
the Hamiltonian%
\begin{equation}
\hat{H}_{\mathrm{XX}} =-\frac{1}{2}\sum_{i=0 } ^{\infty}  \left( \sigma _{i}^{-} \sigma _{i+1}^{+}  + \sigma
_{i}^{-} \sigma _{i-1}^{+} \right) +\frac{h}{2}\sum_{i =0 } ^{\infty} (\sigma _{i}^{z}-\mathbbm{1}),  \label{XX-o}
\end{equation}%
where the summation is over all lattice sites and $h>0$. As usual, $\sigma
_{i}^{\pm }=\left( \sigma _{i}^{x}\pm i\sigma _{i}^{y}\right) /2,$ where $%
\sigma _{i}^{x}$ and $\sigma _{i}^{y}$ together with $\sigma _{i}^{z}$
denote the Pauli spin operators, acting on the $i^{\mathrm{th}}$ spin, and $h$ represents the strength of an
external magnetic field. The commutation relations are%
\begin{equation}
\lbrack \sigma _{i}^{+},\sigma _{k}^{-}]=\sigma _{i}^{z}\delta _{ik},\qquad
\lbrack \sigma _{i}^{z},\sigma _{k}^{\pm }]=\pm 2\sigma _{i}^{\pm }\delta
_{ik}.
\end{equation}%
These operators are nilpotent $(\sigma _{i}^{\pm })^{2}=0$, a property that
will lead to a determinantal form for the correlation functions that we
shall focus on. The other operator satisfies $(\sigma _{i}^{z})^{2}= \mathbbm{1}$.

We do not restrict ourselves to the Hamiltonian $\hat{H}_{\mathrm{XX}}$, and introduce additional interactions in the hopping term. The resulting Hamiltonian is
\begin{equation}
\hat{H}_{\mathrm{Gen}}=-\sum_{i =0 } ^{\infty} \sum_{n\in \mathbb{Z}} a_{n}\left(
\sigma _{i}^{-} \sigma _{i+n}^{+}\right) +\frac{h}{2}\sum_{i=0} ^{\infty} (\sigma
_{i}^{z}-\mathbbm{1}) ,  \label{Hgen-0}
\end{equation}
where the $a_{n}$ denote arbitrary real coefficients which decay at least as 
$a_{n}\sim n^{-1-\epsilon }$ with $\epsilon >0$. 
It has been introduced in \cite{PGT}, and we will analyze it further in sections \ref{app:saddleptsmethod} and \ref{sec:XXextended}. 
Henceforth we work with the Hamiltonian \eref{Hgen-0}.

Let $| \Uparrow \rangle $ denote a ferromagnetic state, which is
characterized by having all the spins up
\begin{equation}
	\vert \Uparrow \rangle = \vert \uparrow, \uparrow , \dots , \uparrow \rangle ,
\end{equation}
satisfying $\sigma_{k}^{+}\vert \Uparrow \rangle =0$ for all $k,$ and the state is
normalized $\langle \Uparrow \mid \Uparrow \rangle =1$. Note that
this state is annihilated by the Hamiltonian, $\hat{H}_{\mathrm{Gen}} \vert \Uparrow
\,\rangle =0$. Then, under the name thermal correlation functions,
defined by : 
\begin{equation}
 F_{j_{1},\dots ,j_{N};l_{1},\dots ,l_{N}}(\beta )=\langle \,\Uparrow
\,| \,{}\sigma _{j_{1}}^{+}\cdots \sigma _{j_{N}}^{+} \ \mathrm{e}^{-\beta 
\hat{H}_{\mathrm{Gen}}} \ \sigma _{l_{1}}^{-}\cdots \sigma _{l_{N}}^{-}{}\,|\,\Uparrow
\,\rangle ,  \label{F}
\end{equation}
the result obtained in \cite{Bog,Bog2,B,BC} (see also \cite{DM}), is the following matrix model representation:
\begin{eqnarray}
\fl F_{j_{1},\dots ,j_{N};l_{1},\dots ,l_{N}}(f,\beta )  =\frac{1}{(2\pi )^{N}N!} 
\int\limits_{\left[ -\pi ,\pi \right] ^{N}} & \dd^{N}\varphi  \prod_{1\leq j<k\leq
N}\left\vert \mathrm{e}^{ \mathrm{i} \varphi _{k}}-\mathrm{e}^{ \mathrm{i} \varphi
_{j}}\right\vert ^{2}\left( \prod_{j=1}^{N}f_{\beta} (e^{\ii \varphi_{j}})\right) \nonumber \\
& \times \hat{s}_{\alpha }\left( e^{ \mathrm{i} \varphi _{1}},\dots ,e^{\mathrm{i} \varphi _{N}}\right) \hat{s}_{\gamma }\left( e^{- \mathrm{i} \varphi _{1}},\dots ,e^{-\mathrm{i} \varphi _{N}}\right) ,
\label{mat}
\end{eqnarray}%
where $\hat{s}_{\alpha }\left( e^{\mathrm{i} \varphi _{1}},\dots
,e^{\mathrm{i} \varphi _{N}}\right) $ is a Schur polynomial, a symmetric polynomial 
\cite{Stanley} whose explicit form is determined by the partition $\alpha $
(which can be conveniently written down in terms of a Young tableaux) and
the weight function $f(e^{\ii \varphi } )$ in the matrix model \eref{mat} is the
generating function of the one-spin flip process ($N=1$ in \eref{F}).  
Therefore, it is given in general by \footnote{Notice that, while there are two indices $j$ and $l$, one is fixed in the difference equation \cite{Bog,Bog2,B,BC} and there is another identical equation with the role of the indices reversed. Both equations satisfy $F_{jl}(0)=\delta _{jl}$. Hence, it is like a one index (random walk) equation, as in \cite{Glauber}, for example.} 
\begin{equation}
f_{ \beta } \left( e^{\ii \varphi } \right) \equiv e^{\beta V(\varphi)} =\sum_{j=-\infty }^{\infty }F_{jl}(\beta
)e^{\ii j \varphi},
\end{equation}%
which, in the case of the XX spin chain reads%
\begin{equation}
f_{\beta} \left( e^{\ii \varphi } \right) = {e}^{\beta (h+\cos \varphi )}.
\label{weight}
\end{equation}%
In random matrix theory, the weight function is also written $f_{\beta} (e^{\ii \varphi } )=\exp{(\beta V( \varphi ) )}$ and $V$ is named the matrix model potential. 
The relationship between the partitions $\alpha $ and $\gamma $ in the
r.h.s. of \eref{mat} and the $j$ and $l$ that index the pattern of flipped
spins in the amplitude \eref{F}, is \cite{B}%
\begin{eqnarray}
\alpha _{r} &=&j_{r}-N+r, \nonumber \\
\gamma _{r} &=&l_{r}-N+r
\end{eqnarray}%
It is clear that \eref{F} is a Loschmidt echo amplitude in imaginary time.
The case of a single-domain wall in the initial state follows by considering
the specific pattern of flipped spins: $j_{r}=N-r$ and $l_{r}=N-r$, in which
case we get in \eref{mat} the void partitions $\alpha =(0,\ldots ,0)$ and $%
\gamma =(0,\ldots ,0)$, which naturally implies \cite{Stanley} that $\hat{s}%
_{\alpha }\left( e^{\mathrm{i} \varphi _{1}},\dots ,e^{\mathrm{i} \varphi _{N}}\right) $ and $\hat{s}_{\gamma }\left( e^{-\mathrm{i} \varphi _{1}},\dots ,e^{-\mathrm{i} \varphi _{N}}\right) $
are both equal to $1$. This clearly corresponds to%
\begin{equation}
\langle ...,\uparrow ,\underset{N}{\underbrace{\downarrow ,...,\downarrow ,\downarrow }}\vert \mathrm{e}^{-\beta \hat{H}_{\mathrm{Gen}}}|\underset{N}{\underbrace{\downarrow ,\downarrow ,...,\downarrow }},\uparrow ,...\rangle ,
\label{equation}
\end{equation}%
and the unitary matrix model description in (\ref{mat}) is without Schur polynomials.
This quantity corresponds to the known Toeplitz determinant representation of Loschmidt echo in imaginary-time
where the initial state is \cite{Echofermion,qreturn}
\begin{equation}
\vert \underbrace{\downarrow ,\downarrow ,...,\downarrow }_{N},\uparrow ,...\rangle  .
\end{equation}

\begin{figure}[hbt]
\centering
\includegraphics[width=0.45\textwidth]{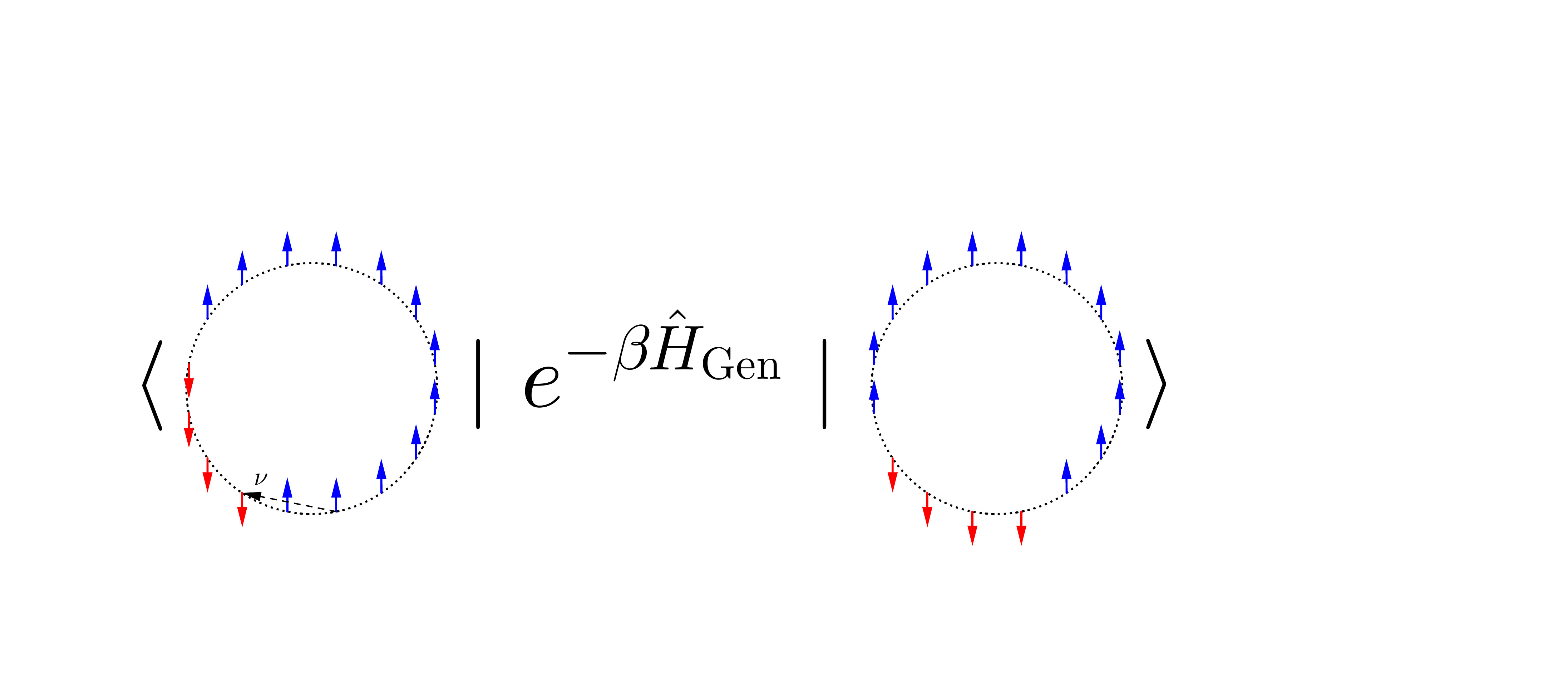}\hspace{0.06\textwidth}
\includegraphics[width=0.45\textwidth]{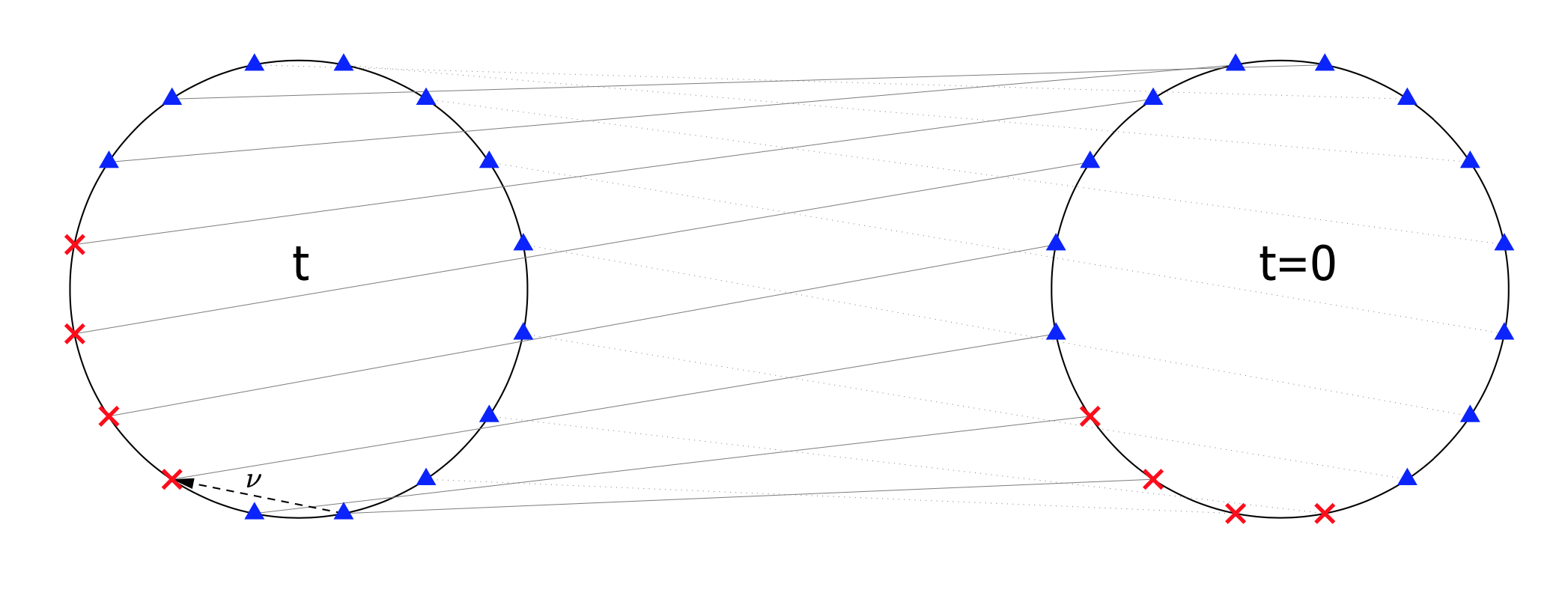}
\caption{Relative shift in the block of flipped spins. Blue is spin up, red is spin down. In this example, $N=4$ and $\nu=2$.}
\label{fig:ringspins}
\end{figure}

Notice that, if one shifts the block of flipped spins by a finite quantity $\nu $, namely considers a state
\begin{equation}
\label{shiftblocknu}
\langle ...,\uparrow ,\underset{N}{\underbrace{\downarrow ,...,\downarrow }},\underset{\nu }{\underbrace{\uparrow ,...,\uparrow }} \vert 
\end{equation}
instead of the ones in \eref{equation},
then one simply has a modification in the weight function of the
matrix model, because in that case, the pattern of flipped spins is $j_{r}=\nu
+N-r $ and the corresponding partition is $\gamma =(\nu ,\ldots ,\nu )$, a
rectangle of $\nu $ columns and $N$ rows, and then $\hat{s}_{\gamma }\left(
e^{\mathrm{i} \varphi _{1}},\dots ,e^{ \mathrm{i} \varphi _{N}}\right) =\prod_{j=1}^{N}e^{\mathrm{i} \nu
\varphi _{j}}$. This is a linear shift in the potential of the matrix model $f_{\beta} (e^{\ii \varphi })=\exp (\beta V( \varphi))$\footnote{Notice that, despite the translational invariance of the problem, there is a term in the matrix model due to the relative shift of the block of flipped spins in the bra with respect to the ket.}.  
Note also that, the variables of the two Schur polynomials in (\ref{mat}) are conjugate and
therefore only relative shifts will appear in the matrix model, with a term $%
\prod_{j=1}^{N}e^{\mathrm{i} (\nu _{b}-\nu _{k})\varphi _{j}}$ and therefore:%
\begin{eqnarray}
\langle ..., \uparrow ,\underset{N}{\underbrace{\downarrow ,...,\downarrow 
}},\underset{\nu _{b}}{\underbrace{\uparrow ,...,\uparrow }} \vert \mathrm{e}%
^{-\beta \hat{H}_{\mathrm{Gen}}}\vert \underset{\nu _{k}}{\underbrace{\uparrow
,...,\uparrow }},\underset{N}{\underbrace{\downarrow ,\downarrow
,...,\downarrow }},\uparrow ,...\rangle \nonumber \\
=\langle ...,\uparrow ,\underset{N}{\underbrace{\downarrow ,...,\downarrow
,\downarrow }} | \mathrm{e}^{-\beta \hat{H}_{\mathrm{Gen}}}|\underset{\nu
_{b}-\nu _{k}}{\underbrace{\uparrow ,...,\uparrow }},\underset{N}{%
\underbrace{\downarrow ,\downarrow ,...,\downarrow }},\uparrow ,...\rangle ,
\end{eqnarray}%
and, as expected from translational invariance, two equal shifts on the two states is equivalent to \eref{equation}. See figure \ref{fig:ringspins}.
In the physical interpretation of \cite{DM}, the linear shift was related to a topological term in gauge theory, along the lines of the notion of momentum polarization \cite{Pol}.

The derivation of the matrix model is not dependent on $\beta $ being real
and thus can be extended to a complex time $w= \beta + \ii t$. Expression \eref{F} can thus be defined for complex time $w \in \mathbb{C}$ as
\begin{equation}
    F_{j_{1},\dots ,j_{N};l_{1},\dots ,l_{N}}( w )=\langle \,\Uparrow \,| \,{}\sigma _{j_{1}}^{+}\cdots \sigma _{j_{N}}^{+} \ \mathrm{e}^{-w \hat{H}_{\mathrm{Gen}}} \ \sigma _{l_{1}}^{-}\cdots \sigma _{l_{N}}^{-}{}\,|\,\Uparrow\,\rangle , 
\end{equation}
for multi-domain wall, and
\begin{equation}
    \mathcal{Z} (w) = \langle ...,\uparrow ,\underset{N}{\underbrace{\downarrow ,...,\downarrow ,\downarrow }}\vert \mathrm{e}^{- w \hat{H}_{\mathrm{Gen}}}|\underset{N}{\underbrace{\downarrow ,\downarrow ,...,\downarrow }},\uparrow ,...\rangle ,
\label{eq:Zamplw}
\end{equation}
for single-domain wall, and the derivation of the matrix model representation continues to hold.

Regarding the equivalent well-known determinantal representation, the Toeplitz and Toeplitz+Hankel
description \cite{Echofermion} corresponds to the unitary matrix model
above and to $Sp(2n)$ matrix integration (see \cite{DM,DGM} and section \ref{app:multidomain}), respectively.
Thus, the choice of the boundary condition determines the symmetry of the
ensemble and the interactions in the chain determine the
specific weight function of the corresponding matrix model. This latter aspect will be discussed below in more detail.

It is also worth mentioning that the analysis above for the case of one
flipped spin $N=1$ leads to the same equation as in Glauber dynamics \cite{Glauber}. In particular, \eref{F} for $N=1$ behaves like the expectation
of a single spin in an infinite ring in \cite{Glauber}, with the time
variable there identified with $\beta $. For this reason, the problem above
is also related to the problems of perfect transfer of information in spin
chains \cite{Bose,Kay}. Generalizations, for example to non-homogeneous
magnetic fields \cite{VZ} (leading for example to the Wannier-Stark problem;
in this case and other generalizations, the corresponding generating
function \eref{weight}, is then the solution of a birth-and-death process)
could be considered here as well and lifted to the $N>1$ setting, described
by the matrix models presented here, which then characterizes the case of
more elusive problem of multi-defect transmissions.

Before turning to the detailed analysis of the matrix model, let us clarify how the present setting is related to a quatum quench. We prepare the initial state $| \psi_0 \rangle = | \Uparrow \rangle$ in an eigenstate of the Hamiltonian of the XX model. Afterwards, we suddenly turn on an additional interaction Hamiltonian (in particular, in Section \ref{sec:XXextended}, we will turn on infinitely many interactions), and calculate the Loschmidt amplitude \eref{F}. This protocol is indeed a quantum quench, see \cite{Heyl:2018abn} for a review. 

\section{Exact results and asymptotics for multi-domain wall configurations}
\label{app:multidomain}

One of the interests of the formulation presented above, is the possibility
of studying amplitudes with initial states which contain several different
domain wall configurations, instead of just one (one single block of flipped
spins on a otherwise completely polarized state). As explained, this extends
the well-known result on Toeplitz and Toeplitz+Hankel determinant
representations \cite{Echofermion} to the setting of minors of such
matrices. Fortunately, the Szeg\H{o} and Fisher-Hartwig asymptotics for the
determinants is extended to the case of minors \cite{Minor,DGM}.

Subsection \ref{sec:defschurs} serves to present useful definitions and properties, as well as to set up the notation, while subsection \ref{sec:toepltizspinchain} contains exact results for a general class of chains.

\subsection{Partitions, Schur and skew-Schur polynomials and minors}
\label{sec:defschurs}

First, we write down basic definitions involving symmetric functions \cite{Stanley}. A partition $\lambda =(\lambda
_{1},\dots ,\lambda _{l})$ is a finite and non-increasing sequence of
positive integers. The number of non-zero entries is named the length of the
partition $l(\lambda )$, and the sum $|\lambda |=\lambda
_{1}+\dots +\lambda _{l(\lambda )}$ is named the weight of the partition.
The entry $\lambda _{j}$ is considered to be zero whenever the index $j$ is
greater than the length of the partition. The notation $(a^{b})$ represents
the partition with $b$ nonzero entries, all equal to $a$. A
partition can be represented as a Young diagram, by placing $\lambda _{j}$
left-justified boxes in the $j$-th row of the diagram. The conjugate
partition $\lambda ^{\prime }$ is obtained as the partition which
diagram has as rows the columns of the diagram of $\lambda $.

Let $f$ be a function on the unit circle,
\begin{equation}
	f (e^{ \mathrm{i} \varphi} ) = \sum_{k \in \mathbb{Z}} d_k e^{ \mathrm{i} k \varphi} ,
\end{equation}
and denote by $T_N (f)$ the $N \times N$ Toeplitz matrix with symbol $f$. This means:
\begin{equation}
	T_N (f) = (d_{j-k} )_{j,k =1} ^{N} .
\end{equation}
We will also use the standard notation $D_N (f)$ to denote the determinant of the Toeplitz matrix $T_N (f)$:
\begin{equation}
	D_N (f) = \det T_N (f) = \det (d_{j-k} )_{j,k =1} ^{N} .
\end{equation}
Besides, $D_N ^{\lambda, \mu} (f)$ will denote the minor of the Toeplitz matrix $T_N (f)$ determined by the two partitions $\lambda, \mu$ (as explained below). When the symbol $f$ is such that the function $\log f$ admits Fourier expansion on the circle, we will write
\begin{equation}
	\log f (e^{ \mathrm{i} \varphi} ) = \sum_{k \in \mathbb{Z}} c_k e^{ \mathrm{i} k \varphi} .
\end{equation}
Notice that, for real-valued $\log f$, the Fourier coefficients satisfy $c_{-k} = c_k ^{\ast}$. 
We now state the celebrated strong Szeg\H{o} limit theorem \cite{Szegoth}.
\begin{theo}[Szeg\H{o} \cite{Szegoth}]Let $f: S^1 \to \mathbb{C}$ be positive and $L_1 (S^1)$, with derivative Holder continuous of positive order. Then 
\begin{equation}
	\lim_{N \to \infty} \frac{ D_N (f) }{ \exp (N c_0) } = \exp \left(  \sum_{k=1} ^{\infty} k  c_k c_{-k} \right) ,
\end{equation}
where $c_k$, for $k=0, \pm 1, \pm 2 , \dots$, is the $k^{\mathrm{th}}$ Fourier coefficient of $\log f$.
\end{theo}

If $x=(x_{1},x_{2},...)$ is a set of variables, the elementary
symmetric polynomials $e_{k}(x)$ and the complete homogeneous polynomials $h_{k}(x)$ are
\begin{eqnarray}
\sum_{k=0}^{\infty }h_{k}(x)z^{k}& =\prod_{j=1}^{\infty }\frac{1}{1-x_{j}z}%
=H(x;z), \\
\sum_{k=0}^{\infty }e_{k}(x)z^{k}& =\prod_{j=1}^{\infty }(1+x_{j}z)=E(x;z).
\end{eqnarray}%
The families $\{h_{k}\}_{k\geq 0}$ and $\{e_{k}\}_{k\geq 0}$ consist of
algebraically independent functions. There are several equivalent ways to define Schur polynomials. The Jacobi-Trudi identities
express Schur polynomials precisely as a Toeplitz minor, generated by the
above functions 
\begin{eqnarray}
s_{\mu }(x)& =\det \left( h_{(j-k+\mu _{k})}(x)\right)
_{j,k=1}^{N}=D_{N}^{\varnothing ,\mu }\left( H(x;z)\right) , \\
s_{\mu ^{\prime }}(x)& =\det \left( e_{(j-k+\mu _{k})}(x)\right)
_{j,k=1}^{N}=D_{N}^{\varnothing ,\mu }\left( E(x;z)\right) ,
\end{eqnarray}%
where $l(\mu ),l(\mu ^{\prime })\leq N$, respectively, and $\varnothing $
denotes the empty partition. The skew-Schur polynomials are
\begin{equation}
s_{\mu /\lambda }(x)=D_{N}^{\lambda ,\mu }(H(x;z)),\qquad s_{(\mu /\lambda )^{\prime }}(x)=D_{N}^{\lambda ,\mu }(E(x;z)),  \label{skewschur}
\end{equation}%
where $l(\mu ),l(\mu ^{\prime })\leq N$ respectively. These polynomials
vanish if $\lambda \nsubseteq \mu $. In the expressions above,
\begin{equation}
    D_{N}^{\lambda,\mu}(f)=\det{T_{N}^{\lambda,\mu}(f)},
\end{equation}
where
\begin{equation}
	T_{N}^{\lambda,\mu}(f)=(d_{j-\lambda_{j}-k+\mu_{k})})_{j,k=1}^{N},
\end{equation}
and $\lambda$ and $\mu$ are integer partitions that can be shown to describe a specific striking of rows and columns of a larger Toeplitz matrix, see \cite{Minor} for details.

The Andreief identity, which establishes for example the equivalence between the Toeplitz determinant and a unitary matrix model \cite{RM} (a relationship also applicable to Toeplitz+Hankel determinants and matrix integration over other classical Lie groups \cite{DGM}), also comprises an analogous identity for the more general setting of minors discussed here. In particular, it holds that \cite{Minor}
\begin{eqnarray}
\label{heinemin}
\fl D_{N}^{\lambda,\mu}(f) &= \int_{U(N)}  \overline{s_{\lambda}(M)}s_{\mu}(M)f(M) dM  \\
\fl	& = \frac{1}{N!} \int_{[0, 2 \pi]^N} \frac{\mathrm{d}^N \varphi }{(2\pi)^N}s_{\lambda}(e^{- \mathrm{i} \varphi_{1}},...,e^{- \mathrm{i} \varphi_{N}})s_{\mu}(e^{ \mathrm{i} \varphi_{1}},...,e^{ \mathrm{i} \varphi_{N}})\prod_{j=1}^{N}f(e^{ \mathrm{i} \varphi_{j}})\prod_{1\leq j<k\leq N}|e^{ \mathrm{i} \varphi_{j}}-e^{ \mathrm{i} \varphi_{k}}|^{2} , \nonumber
\end{eqnarray}
where $s_{\lambda},s_{\mu}$ are Schur polynomials.

\subsection{Asymptotics of Toeplitz minors and spin chain interpretations}
\label{sec:toepltizspinchain}

For the case of large $N$, the determinant contribution and the
combinatorial one, depending on the partitions $\lambda $ and $\mu $ appear
in a completely factorized way and can be treated independently \cite%
{Minor,DGM}. Let us denote \eref{mat} by $D_{N}^{\lambda ,\mu }\left(
f\right) $ and, the corresponding determinant, which is the same matrix model
but without partitions, which describes the amplitude \eref{equation}, by $%
D_{N}\left( f\right) $. Then, it holds, for symbols $f(e^{ \mathrm{i} \theta })$ in the
Szeg\H{o} class and also with FH singularities. Then, as $N\rightarrow
\infty $, 
\begin{equation}
D_{N}^{\lambda ,\mu }(f)\sim D_{N}(f)\sum_{\nu }s_{\lambda /\nu }(y)s_{\mu
/\nu }(x),  \label{mainsum}
\end{equation}%
where the variables $y,x$ are such that $f(z)=H(y;z^{-1})H(x;z)$, and the
sum runs over all partitions $\nu $ contained in $\lambda $ and $\mu $. In 
\cite{Minor}, Bump and Diaconis originally obtained other explicit expressions for 
\eref{mainsum}, involving Laguerre polynomials and simplifying considerably
in the case of only one non-trivial partition. Let us simply present an explicit
example. Again, for a generic symbol
\begin{equation}
f \left( e^{\mathrm{i} \varphi }\right) = \sum_{k\in \mathbb{Z} }d_{k}e^{ \mathrm{i} k \varphi }=\exp \left( \sum_{k \in \mathbb{Z} } c_{k}e^{ \mathrm{i} k \varphi } \right) ,
\end{equation}
corresponding to a choice of Hamiltonian as in \eref{Hgen-0}, and for example two partitions in the antisymmetric representation $\lambda =\mu =(1^{2})$ 
\begin{equation}
\frac{D_{N}^{\lambda ,\mu }\left( f\right) }{D_{N}\left( f\right) }\sim 
\frac{1}{4}c_{1}^{2}c_{-1}^{2}+c_{1}c_{-1}-\frac{1}{2}c_{-2}c_{1}^{2}-\frac{1%
}{2}c_{-1}^{2}c_{2}+c_{2}c_{-2}+1. \label{pol}
\end{equation}%
Notice that the asymptotics of the determinant in the denominator is $%
D_{N}\left( f\right) \sim \exp \left( \sum_{k=1}^{\infty
}kc_{k}c_{-k}\right) .$ This translates to the following explicit Loschmidt
echo amplitude as follows:%
\begin{eqnarray}
\fl \langle ...,\uparrow ,\uparrow ,\downarrow ,\downarrow ,\uparrow ,\underset{%
N-2}{\underbrace{\downarrow ,...,\downarrow ,\downarrow }}\vert \mathrm{e}%
^{-\beta \hat{H}_{\mathrm{Gen}}}\vert \underset{N-2}{\underbrace{\downarrow ,\downarrow
,...,\downarrow }},\uparrow ,\downarrow ,\downarrow ,\uparrow ,\uparrow
,...\rangle \nonumber \\ \sim P(c_{\pm 1},c_{\pm 2})\langle ...,\uparrow ,\underset{N}{%
\underbrace{\downarrow ,...,\downarrow ,\downarrow }}\vert \mathrm{e}%
^{-\beta \hat{H}_{\mathrm{Gen}} }\vert \underset{N}{\underbrace{\downarrow ,\downarrow
,...,\downarrow }},\uparrow ,...\rangle ,
\end{eqnarray}%
where $P(c_{\pm 1},c_{\pm 2})$ is the polynomial on the right hand side of (\ref{pol}). Taking $c_{1}=c_{-1}=w$ while $c_{k}=c_{-k}=0$ for $k>1$ then gives the
corresponding amplitude for the XX spin chain%
\begin{equation}
D_{N}^{\lambda ,\mu }\left( f\right) \sim \left( \frac{1}{4} w^{4}+w^{2}+1\right) \exp \left( w^{2}\right) .
\end{equation}

\section{Phase transitions with saddle-point method}
\label{app:saddleptsmethod}

The present section is dedicated to the analysis of the Loschmidt amplitudes at large $N$, through the study of large $N$ asymptotics of the equivalent matrix model description, in a scaling limit. 
The content of this section is presented in a general formalism, recovering the cases of interest for the rest of the paper as special instances.

Consider the generic spin Hamiltonian \eref{Hgen-0}.
We use the result of \cite{PGT}, and write the Loschmidt amplitude at imaginary time
\begin{equation}
    \mathcal{Z} (\beta ) = \langle ...,\uparrow ,{\underbrace{\downarrow ,...,\downarrow ,\downarrow }_{N}}\vert \mathrm{e}^{- \beta \hat{H}_{\mathrm{Gen}}}|{\underbrace{\downarrow ,\downarrow ,...,\downarrow }_{N}},\uparrow ,...\rangle ,
\label{eq:ZamplHgen}
\end{equation}
as the partition function of the matrix model
\begin{equation}
\mathcal{Z} (\beta )= \frac{1}{N!}\int_{-\pi} ^{\pi} \frac{ \mathrm{d}
\varphi_1}{2 \pi} \cdots \int_{-\pi} ^{\pi} \frac{ \mathrm{d} \varphi_N}{2
\pi} \prod_{1\le j<k \le N} \left| e^{\mathrm{i} \varphi_j} - e^{\mathrm{i}
\varphi_k} \right|^2 e^{\beta \sum_{j=1} ^N V ( \varphi_j ) } ,
\label{Zgen}
\end{equation}
with potential
\begin{equation}
V(\varphi )= h + 2 \sum_{n=1} ^{\infty} \left[ \frac{g_n}{n} \cos (n \varphi) + \frac{\tilde{g}_n}{n} \sin (n \varphi) \right] .  \label{Vgen}
\end{equation}
The coefficients of $V (\varphi)$ and these of the Hamiltonian $\hat{H}_{\mathrm{Gen}}$ in \eref{Hgen-0} are related through \cite{PGT}
\begin{equation}
\label{eq:coefsVvsH}
	g_n = n ( a_n + a_{-n}) , \quad 	\tilde{g}_n = \mathrm{i} n ( a_n - a_{-n}) .
\end{equation}
Setting $g_n= e^{- n \alpha}$ and $\tilde{g}_n=0$ we recover the matrix
model \eref{Zw}, studied later in section \ref{sec:interactionBaik}, as a particular case. Choosing instead $g_1 = 1$, $\tilde{g}_1 = - \mathrm{i} v $ 
and all the other vanishing, $g_n = 0 = \tilde{g}_n$ for $n>1$, corresponds to the Hamiltonian \eref{HhopGWv} considered in section \ref{sec:GWasym}.\par

In subsection \ref{sec:HGenspectrum} we discuss properties of the Hamiltonian $\hat{H}_{\mathrm{Gen}}$ defined in \eref{Hgen-0}, and in the rest of the section we present the study of large $N$ phase transition in the matrix ensemble \eref{Zgen}.

\subsection{Spin Hamiltonian with interaction beyond nearest neighbour}
\label{sec:HGenspectrum}

To analyze the properties of the Hamiltonian \eref{Hgen-0} and the dispersion relation, we take a finite chain with $L+1$ sites\footnote{Eventually, we take the limit $L \to \infty$. Chains of finite length will be analyzed in section \ref{sec:finite}.}, with periodic boundary conditions, $L+1 \sim 0$. For simplicity we assume $L+1$ is odd, so that each spin has $L/2$ neighbours on its left and the same amount on its right. The finite length version of the Hamiltonian \eref{Hgen-0} reads:
\begin{equation}
\hat{H}_{\mathrm{Gen}}=-\sum_{i =0 } ^{L} \sum_{n=1} ^{L/2} \left( a_{n}
\sigma _{i}^{-} \sigma _{i+n}^{+} + a_{-n} \sigma _{i}^{-} \sigma _{i-n}^{+} \right) +\frac{h^{\prime} }{2}\sum_{i=0} ^{L} (\sigma
_{i}^{z}-\mathbbm{1}) ,  \label{Hgen-finiteL}
\end{equation}
where we reabsorbed $a_0$ into $h^{\prime} = h+ a_0$ (we henceforth drop the $^{\prime}$). 
We note that
\begin{equation}
\label{HgencommSZtot}
	\left[ \hat{H}_{\mathrm{Gen}}, \hat{S}^{z} \right]=0 , \qquad \text{ where } \ \hat{S}^{z} = \sum_{i=0} ^{L} \sigma_i ^{z} ,
\end{equation}
meaning that the total number of spins down is a conserved quantity. Denoting $\mathcal{H}$ the Hilbert space of the theory, we can conveniently separate $\mathcal{H}$ into
\begin{equation}
	\mathcal{H} = \bigoplus_{r= 0} ^{L} \mathcal{H}_r ,
\label{separateHilbert}
\end{equation}
where $\mathcal{H}_r$ denotes the space of states with exactly $r$ spins down, all the other spins up. The commutation relation \eref{HgencommSZtot} guarantees that different sectors $\mathcal{H}_r$ do not mix under the action of $\hat{H}_{\mathrm{Gen}}$. The space $\mathcal{H}_0$ is spanned by the vacuum $\vert \Uparrow \rangle $, and $\mathcal{H}_1$ is spanned by states with a single spin down, which throughout this subsection we denote by
\begin{equation}
	\vert \downarrow_k \rangle \equiv \vert \uparrow , \uparrow , \dots , \underset{k^{\mathrm{th}}}{\downarrow }, \uparrow , \dots \rangle , \qquad k= 0, \dots, L .
\end{equation}
The action of the Hamiltonian \eref{Hgen-finiteL} on such states is
\begin{eqnarray}
\fl	\hat{H}_{\mathrm{Gen}} \vert \downarrow_k \rangle &=& - \sum_{i=0} ^{L} \left[ \sum_{n =1} ^{L/2} \left( a_n \sigma_{i} ^{-} \sigma^{+} _{i+n} + a_{-n} \sigma_{i} ^{-} \sigma^{+} _{i-n} \right)   - \frac{h}{2} \left(  \sigma_i ^{z} - \mathbbm{1} \right) \right]  \vert \downarrow_k \rangle \nonumber \\ 
\fl	&=& - \sum_{n=1} ^{L/2} \big(   a_{n} \vert \downarrow_{k-n} \rangle + a_{-n} \vert \downarrow_{k+n} \rangle \big) - h  \vert \downarrow_k \rangle .
\end{eqnarray}
The effect of the magnetic field $h$ on states in $\mathcal{H}_r$ is a contribution $-rh$, for every $r$.

We let $s=0, \dots, L$ be a discrete variable running on the dual lattice and introduce
\begin{equation}
\label{eq:discrangvars2}
	\varphi_s = \frac{2 \pi }{L+1} \left( s - \frac{L}{2} \right) .
\end{equation}
For the eigenstates of $\hat{H}_{\mathrm{Gen}}$ in $\mathcal{H}_1$ we use the standard ansatz
\begin{equation}
\label{1Pansatz}
	\vert s \rangle = \frac{1}{\sqrt{L+1}} \sum_{k=0}^{L} e^{\ii \varphi_s k} \vert \downarrow_k \rangle ,
\end{equation}
compatible with translation invariance, and obtain
\begin{eqnarray}
\fl \hat{H}_{\mathrm{Gen}}  \vert s \rangle & = & \frac{1}{\sqrt{L+1}}  \sum_{k=0} ^{L} e^{\ii \varphi_s k} \left[ - \sum_{n=1} ^{L/2} \left( a_{n} \vert \downarrow_{k-n} \rangle + a_{-n} \vert \downarrow_{k+n} \rangle \right)  - h  \vert \downarrow_k \rangle \right] \nonumber \\ 
\fl	& = & \frac{1}{\sqrt{L+1}}  \sum_{k^{\prime}=0} ^{L} e^{\ii \varphi_s k^{\prime}} \left[ - \sum_{n=1}^{L/2}\left( a_n e^{\ii \varphi_s n} + a_{-n} e^{- \ii \varphi_s n}  \right) - h \right]  \vert \downarrow_{k^{\prime}} \rangle  .
\end{eqnarray}
Hence the states $\vert s \rangle$ satisfy the eigenvalue equation $\hat{H}_{\mathrm{Gen}}  \vert s \rangle = \varepsilon_s  \vert s \rangle $ with energy
\begin{equation}
\label{1Penergy}
	\varepsilon_s = -h - \sum_{n=1} ^{L/2} \left[ (a_n + a_{-n} ) \cos (\varphi_s n ) + \ii (a_n - a_{-n} ) \sin (\varphi_s n) \right] .
\end{equation}
Setting $a_1 = a_{-1} = \frac{1}{2}$ and $a_n = a_{-n} = 0 $ for $n \ne 1$ we recover the well known dispersion relation of the XX chain.\par
The presence of additional interactions, compared to the XX chain, does not allow to reconstruct the full spectrum knowing the eigenstates in $\mathcal{H}_1$. At this point, it is worthwhile to compare the present setting with that discussed in \cite[Appendix D]{Echofermion}, where a Hamiltonian with general dispersion relation (gdr) was introduced. In our notation it reads
\begin{equation}
	\hat{H}_{\mathrm{gdr}} = - \sum_{i=0} ^{L} \sum_{n=1} ^{L/2} \left(  a_n \hat{c}^{\dagger} _{j} \hat{c}_{j+n}  + a_{-n} \hat{c}^{\dagger} _{j} \hat{c}_{j-n}  \right) , \label{Hgen-gdr}
\end{equation}
with $\hat{c}^{\dagger} _j, \hat{c}_j$ fermionic creation and annihilation operators. This Hamiltonian describes by construction a system of free fermions. Under the action of the Hamiltonian $\hat{H}_{\mathrm{gdr}}$ the Hilbert space $\mathcal{H}$ of the theory is separated as in \eref{separateHilbert}, but now $r$ counts the number of fermionic excitations. The dispersion relation of the Hamiltonian \eref{Hgen-gdr} is again given by \eref{1Penergy} \cite{Echofermion}, because the two Hamiltonians $\hat{H}_{\mathrm{Gen}}$ and $\hat{H}_{\mathrm{gdr}}$ have the same eigenfunctions with the same corresponding eigenvalues in the subspace $\mathcal{H}_1 \subset \mathcal{H}$. However, this statement is not true in $\mathcal{H}_{r \ge 2}$. 
Indeed, the Hamiltonian \eref{Hgen-gdr} restricted on $\mathcal{H}_r$ describes $r$ free fermions, and its eigenfunctions in $\mathcal{H}_r$ are totally anti-symmetric products of $r$ single-particle eigenfunctions. 
These functions are not eigenfunctions of the Hamiltonian \eref{Hgen-finiteL}, as can be easily checked already at $r=2$. Moreover, the eigenfunctions of the generic Hamiltonian \eref{Hgen-finiteL} on $\mathcal{H}_2$ are not found by direct application of the coordinate Bethe ansatz (see for example \cite{Franchini2016} for an introduction to the subject) unless $a_{\pm n} = 0$ for all $n>1$, suggesting that interactions beyond nearest neighbour might spoil integrability. 

With these considerations in mind, we conclude that the Loschmidt amplitudes $\mathcal{Z} (\beta )$ associated to the Hamiltonian \eref{Hgen-gdr} of \cite[Appendix D]{Echofermion} are equal to the ones studied in the present work, defined in \eref{eq:ZamplHgen} using the Hamiltonian \eref{Hgen-finiteL}. This is because the matrix model representation of the amplitude $\mathcal{Z} (\beta)$ arises as a Toeplitz determinant of amplitudes computed in $\mathcal{H}_1$, which is the subspace of the Hilbert space $\mathcal{H}$ on which the restrictions of the two Hamiltonians \eref{Hgen-finiteL} and \eref{Hgen-gdr} coincide. In other words, although the two Hamiltonians are different, they lead to the same Loschmidt amplitudes \eref{eq:ZamplHgen}-\eref{Zgen}, as these observables only depend on the $\mathcal{H}_1$ subsector of the Hilbert space of the theory. 
We also remark that the matrix model representation is obtained in \cite{PGT} with a procedure that does not rely on the free fermion formalism.

\subsection{Set up: large $N$ limit and eigenvalue density}
\label{sec:setuplargeN}

We are interested in the scaling limit $N \to \infty$ with 
\begin{equation}
\beta/ N \equiv \gamma \ \mathrm{fixed} .
\end{equation}
of the Loschmidt amplitude $\mathcal{Z} (\beta )$ of \eref{eq:ZamplHgen}-\eref{Zgen}. As a first step, we rewrite the partition function \eref{Zgen} as: 
\begin{equation}
\mathcal{Z} (\beta )= \frac{e^{\beta h N }}{N!}\int_{-\pi} ^{\pi} \frac{ \mathrm{d}
\varphi_1}{2 \pi} \cdots \int_{-\pi} ^{\pi} \frac{ \mathrm{d} \varphi_N}{2
\pi} e^{ N^2 S ( \varphi ) } ,
\end{equation}
with action 
\begin{eqnarray}
 S (\varphi) &= \frac{ 2 \beta}{N^2} \sum_{j=1} ^N \sum_{n=1} ^{\infty} \left[ 
\frac{g_n}{n} \cos (n \varphi_j) + \frac{\tilde{g}_n}{n} \sin (n \varphi_j) %
\right] \nonumber \\
&+ \frac{1}{N^2}\sum_{1 \le j < k \le N} \log \left( 2 \sin \left( 
\frac{ \varphi_j - \varphi_j}{2} \right) \right)^2 .
\end{eqnarray}
Taking the large $N$ limit, the leading contribution to the partition
function \eref{Zgen} comes from the saddle points of the action $S (\varphi)
$. In this limit, we define $x= j/N$ and replace the sums by integrals, so
that the action becomes 
\begin{eqnarray}
 \lim_{N \to \infty} S ( \varphi) & = 2 \gamma \sum_{n=1} ^{\infty} \int_0 ^1
\dd x \left[ \frac{g_n}{n} \cos (n \varphi (x)) + \frac{\tilde{g}_n}{n} \sin (n
\varphi (x) ) \right] \\
& + \int_{0} ^1 \dd x \int_0 ^1 \dd x^{\prime} \log \left\vert 2
\sin \left( \frac{ \varphi (x) - \varphi (x^{\prime}) }{2} \right) \right\vert , \nonumber 
\end{eqnarray}
and its saddle points are determined by the functional equation $\frac{
\delta S }{\delta \varphi } = 0$. Thus we pursue a function $\varphi : [0,1] \to [-
\pi, \pi]$ which solves the saddle point equation 
\begin{eqnarray}
\fl 2 \gamma \sum_{n=1} ^{\infty} \left[ - g_n \sin (n \varphi (x)) + \tilde{g}_n \cos (n \varphi (x) ) \right] + \mathrm{P} \int_{0} ^1 \dd x^{\prime} \cot
\left( \frac{ \varphi (x) - \varphi (x^{\prime}) }{2} \right) =0 ,
\end{eqnarray}
for all $x \in [0,1]$. We can restate the problem as follows. Consider the
eigenvalue density 
\begin{equation}
\rho (\varphi) = \frac{ \mathrm{d} x}{\mathrm{d} \varphi} ,
\end{equation}
which is normalized: 
\begin{equation}
\int \rho (\varphi) \mathrm{d} \varphi = 1 .  \label{normalizationrho}
\end{equation}
Then the saddle point equation is rewritten as a singular integral equation: 
\begin{eqnarray}
 2 \gamma \sum_{n=1} ^{\infty} \left[ - g_n \sin (n \varphi ) + \tilde{g}_n
\cos (n \varphi ) \right] + \mathrm{P} \int \mathrm{d} \vartheta \rho
(\vartheta) \cot \left( \frac{ \varphi - \vartheta }{2} \right) = 0 ,
\label{SPErho}
\end{eqnarray}
for all $\varphi \in \mathrm{supp} \rho$, where $\mathrm{P} \int $ means the
principal value of the integral. Therefore, we look for a function $\rho$,
with $\mathrm{supp} \rho \subseteq [- \pi, \pi]$, which solves the saddle
point equation \eref{SPErho} and the normalization condition %
\eref{normalizationrho}.

\subsection{Solution at small $\protect\gamma$}
\label{app:smallgammasol}

We start looking for an eigenvalue density supported on the whole circle, $%
\mathrm{supp} \rho = [- \pi, \pi]$. To solve Eq. \eref{SPErho} we use the
identity 
\begin{equation}
\fl \mathrm{P} \int_{-\pi} ^{\pi} \mathrm{d} \vartheta \rho (\vartheta) \cot
\left( \frac{ \varphi - \vartheta }{2} \right) = 2 \sum_{n=1} ^{\infty}
\int_{-\pi} ^{\pi} \mathrm{d} \vartheta \rho (\vartheta) \left[ \sin (n
\varphi) \cos (n \vartheta) - \cos (n \varphi) \sin (n \vartheta) \right] ,
\end{equation}
and expand $\rho (\vartheta)$ in Fourier series: 
\begin{equation}
\rho (\vartheta) = a_0 + \sum_{n=1} ^{\infty} \left[ a_n \cos (n \vartheta)
+ b_n \sin (n \vartheta) \right] ,
\end{equation}
with 
\begin{equation}
a_n = \frac{1}{\pi} \int_{-\pi} ^{\pi} \mathrm{d} \vartheta \rho (\vartheta)
\cos ( n \vartheta) , \quad b_n = \frac{1}{\pi} \int_{-\pi} ^{\pi} \mathrm{d}
\vartheta \rho (\vartheta) \sin ( n \vartheta) ,
\end{equation}
for all $n=0,1,\dots$. Eq \eref{SPErho} can then be rewritten as: 
\begin{equation}
2 \gamma \sum_{n=1} ^{\infty} \left[ g_n \sin (n \varphi ) - \tilde{g}_n
\cos (n \varphi ) \right] = 2 \pi \sum_{n=1} ^{\infty} \left[ a_n \sin (n
\varphi) - b_n \cos (n \varphi) \right] ,
\end{equation}
and from the orthogonality of the Fourier basis it immediately follows that: 
\begin{equation}
a_n = \frac{ \gamma}{\pi} g_n , \quad b_n = \frac{ \gamma}{\pi} \tilde{g}_n ,
\end{equation}
for all $n =1 , 2 ,\dots$. Besides, the normalization condition %
\eref{normalizationrho} fixes $a_0 = \frac{1}{2 \pi}$. Therefore, under the
requirement $\mathrm{supp} \rho = [- \pi, \pi]$, we obtained the eigenvalue
density 
\begin{equation}
\rho (\varphi )=\frac{1}{2\pi }\left[ 1+2\gamma \sum_{n=1}^{\infty } \left(
g_{n} \cos (n \varphi ) + \tilde{g}_{n} \sin (n \varphi ) \right) \right] ,
\label{rhoPhI}
\end{equation}
which is a valid solution as long as it defines a probability measure. In
particular, from the fact that, on the circle, $\vert \varphi_j - \varphi_k \vert \le 2 \pi$, it follows that $\rho (\varphi) = \frac{ \mathrm{d} x }{ \mathrm{d} \varphi}$ must be non-negative defined: 
\begin{equation}
\rho (\varphi) \ge 0 , \qquad -\pi < \varphi \le \pi .
\end{equation}
If 
\begin{equation}
\min_{- \pi <\varphi \le \pi } \sum_{n=1}^{\infty } \left( g_{n} \cos (n
\varphi ) + \tilde{g}_{n} \sin (n \varphi ) \right) < 0 ,
\end{equation}
the non-negativity condition imposes an upper bound to the parameter $\gamma$, hence the solution \eref{rhoPhI} is valid until $\gamma \le \gamma_c$,
with critical value: 
\begin{equation}
\gamma_c = \left[ - 2 \min_{- \pi <\varphi \le \pi } \sum_{n=1}^{\infty }
\left( g_{n} \cos (n \varphi ) + \tilde{g}_{n} \sin (n \varphi ) \right) %
\right]^{-1} > 0.
\end{equation}
We now use the the solution \eref{rhoPhI} in the weak coupling phase to prove the following statement.

\begin{prop}
\label{propszego}
Consider a unitary matrix model as in \eref{Zgen}. Then, in the limit $N \to \infty$ with $\beta/N \equiv \gamma$ fixed, for all $0 \le \gamma \le \gamma_c$ the free energy $\mathcal{F} = \log 
\mathcal{Z} (\beta) /N^2 $ coincides with the result of Szeg\H{o}'s theorem.
\end{prop}

\textit{Proof.} It follows from direct calculation of the derivative of $\mathcal{F}$. In the large $N$ limit:
\begin{eqnarray*}
\fl \frac{ \dd \mathcal{F} }{ \dd \gamma} ( \gamma \le \gamma_c) & = \int_{-\pi} ^{\pi} \dd \varphi  \rho (\varphi) V (\varphi) \nonumber \\
\fl	& =  \int_{-\pi} ^{\pi} \frac{\dd \varphi}{2 \pi}  \left[ 1 + 2 \gamma \sum_{n=1} ^{\infty} \left( g_n \cos (n \varphi ) + \tilde{g}_n \sin (n \varphi) \right) \right] \nonumber \\
\fl & \times \left[ h + 2 \sum_{m=1}^{\infty} \left( \frac{ g_m }{m} \cos (m \varphi) +  \frac{ \tilde{g}_m }{m} \sin (m \varphi) \right) \right] \nonumber \\
\fl	& = h + \frac{ 2 \gamma }{\pi} \sum_{n=1} ^{\infty} \sum_{m=1} ^{\infty}  \int_{-\pi} ^{\pi} \dd \varphi  \left[ \frac{ g_n g_m }{m} \cos (n \varphi) \cos (m \varphi) +  \frac{ \tilde{g}_n \tilde{g}_m }{m} \sin (n \varphi) \sin (m \varphi) \right] \nonumber \\
\fl	&= h + 2 \gamma \sum_{n=1} ^{\infty} \frac{ g_n ^2 + \tilde{g}_n ^2 }{n} ,
\end{eqnarray*}
where in the last two equalities we have used the orthogonality relations. The integration with boundary condition $\mathcal{Z} (0) =1$ \footnote{This follows directly from $\mathcal{Z} (0) = \int_{U(N)} dU$, where $dU$ is the normalized Haar measure on $U(N)$.} immediately gives:
\begin{equation}
\mathcal{F}  ( \gamma \le \gamma_c) = \gamma h + \gamma^2 \sum_{n=1} ^{\infty} \frac{ g_n ^2 + \tilde{g}_n ^2 }{n} ,
\label{Fweakgen}
\end{equation}
matching the result of direct application of Szeg\H{o}'s theorem, up to trivially scaling $\beta/N=\gamma$.

\subsection{Specializations at small $\protect\gamma$}

We now specialize the result \eref{rhoPhI} to some fundamental examples, which will be of interest in sections \ref{sec:GWasym} and \ref{sec:XXextended}.

As a first case, we consider the original XX spin chain, with only nearest-neighbour
interaction. We set $g_1=1 $ and all the other coefficients to zero: $%
g_{n>1} = \tilde{g}_{n\ge 1} =0$. This choice gives the Gross-Witten-Wadia
matrix model \cite{GW,Wadia}. Then, the eigenvalue density at small $\gamma$
is given by 
\begin{equation}
\rho_{\mathrm{GW}} (\varphi) = \frac{1}{2 \pi} \left[ 1 + 2 \gamma \cos
(\varphi) \right] ,
\end{equation}
with critical value $\gamma_c=\frac{1}{2}$. From \eref{Fweakgen}, we have
that the free energy takes the simple expression $\mathcal{F} ( \gamma \le
\gamma_c) = \gamma h + \gamma^2$.

Another case of relevance is the spin chain with exponentially decaying
interaction, corresponding to the choice $g_n = e^{- n \alpha}$ and $\tilde{g%
}_n =0$, leading to the model \eref{Zw}. In this case the potential is
logarithmic: 
\begin{equation}
V (\varphi) = h - \log (1- e^{- \alpha} e^{ \mathrm{i} \varphi} )(1-
e^{-\alpha} e^{ - \mathrm{i} \varphi} ) ,  \label{VBaik}
\end{equation}
and the eigenvalue density in the small $\gamma$ phase, $0 \le \gamma \le \gamma_c$, is: 
\begin{equation}
\rho_{\mathrm{B}} (\varphi) = \frac{1}{2 \pi} \left[ 1 + 2 \gamma \sum_{n=1}
^{\infty} e^{-n \alpha} \cos (n \varphi) \right] .
\end{equation}
The series in latter expression can be computed explicitly, so that the
eigenvalue density takes the form: 
\begin{equation}
\rho_{\mathrm{B}} (\varphi) = \frac{1}{2 \pi} \left[ 1 + \gamma e^{-
\alpha} \left( \frac{ \cos (\varphi) - e^{-\alpha} }{ 1 + e^{- 2 \alpha} - 2
e^{- \alpha} \cos (\varphi) }\right) \right] .  \label{rhoIBaik}
\end{equation}
There are (at least) two ways to see this. One could expand $2\cos (n \varphi) = (e^{%
\mathrm{i} \varphi})^n + (e^{- \mathrm{i} \varphi})^n$ and obtain the
geometric series with argument $e^{- \alpha \pm \mathrm{i} \varphi}$.
Otherwise, one could use the relation $\cos (n \varphi) = T_n ( \cos (
\varphi))$, where $T_n$ in the $n$-th Chebyshev polynomial of first kind,
and recognize the generating function of the Chebyshev polynomials with
argument $e^{-\alpha}$. The minimum of expression \eref{rhoIBaik} is
located at $\varphi=\pm \pi$, whence the critical value is: 
\begin{equation}
\gamma_c = \frac{ 1 +e^{- \alpha} }{2 e^{- \alpha}} .
\end{equation}
The same result was obtained by Baik in \cite{JB} adopting different and more powerful techniques.
We can also get the free energy from formula \eref{Fweakgen}: 
\begin{equation}
\mathcal{F}_{\mathrm{B}} ( \gamma \le \gamma_c) =  \gamma h + \gamma^2 \sum_{n=1}
^{\infty} \frac{e^{-2 \alpha}}{n} =  \gamma h - \gamma^2 \log (1-e^{-2 \alpha}).
\label{FBaikweak}
\end{equation}

As a third example, we consider the potential: 
\begin{equation}
V (\varphi) = h + 2 \sum_{n=1} ^{\infty} \frac{ e^{- n \alpha} }{n^{1+p}} \cos
(n \varphi) = h + \mathrm{Li}_{1+p} ( e^{- \alpha + \mathrm{i} \varphi}) + 
\mathrm{Li}_{1+p} ( e^{- \alpha - \mathrm{i} \varphi}) ,  \label{Vpolylog}
\end{equation}
with $p \in \mathbb{Z}$ and $\alpha >0$ if $p \le 0$ and $\alpha \ge 0$ if $%
p >0$. For $\alpha > 0$ this is a modification of the spin chain with
exponentially decaying interaction, however, for $p\in \mathbb{Z}_{>0}
$ we can set $\alpha=0$ and obtain a spin chain with long range interaction $\sim 1/n^{1+p}$. 
The eigenvalue density at small $\gamma$ is obtained
plugging $g_n = e^{- n \alpha}/n^p$ and $\tilde{g}_n=0$ in the general
solution \eref{rhoPhI}. The result is: 
\begin{equation}
\rho (\varphi) = \frac{1}{2 \pi} \left[ 1 + 2 \gamma \left( \mathrm{Li}_{p}
(e^{- \alpha + \mathrm{i} \varphi }) + \mathrm{Li}_{p} (e^{- \alpha - 
\mathrm{i} \varphi }) \right) \right] ,
\end{equation}
whose minimum is located at $\varphi = \pm \pi$, thus the critical value is
given by: 
\begin{equation}
\gamma_c = - \left( 2 \mathrm{Li}_p (- e^{-\alpha}) \right)^{-1} .
\end{equation}
The solution \eref{rhoIBaik} is recovered as a special case when $p=0$.
Furthermore, for the long range interaction $p\in \mathbb{Z}_{>0}$ and $\alpha=0$,
the critical value is 
\begin{equation}
\gamma_c = \frac{1}{2(1-2^{1-p}) \zeta (p) } ,
\end{equation}
where $\zeta (p)$ is the Riemann zeta function, and we have used the
relation between the polylogarithm of order $p$ at $-1$ and $\zeta (p)$.
From \eref{Fweakgen}, we have that the free energy is: 
\begin{equation}
\mathcal{F} ( \gamma \le \gamma_c) =  \gamma h + \gamma^2 \sum_{n=1} ^{\infty} \frac{%
e^{-2 \alpha}}{n^{2p+1}} =  \gamma h + \gamma^2 \mathrm{Li}_{2p+1} (e^{- 2 \alpha}) .
\end{equation}
Again, for $p=0$ we recover the previous result \eref{FBaikweak}. For the
long range interaction $\alpha=0$ instead we have: 
\begin{equation}
\mathcal{F} ( \gamma \le \gamma_c) = \gamma h + \gamma^2 \zeta (2p+1) .
\end{equation}

\subsection{Solution at large $\protect\gamma$}

When the coupling $\gamma$ exceeds the value for which the eigenvalue density obtained in Eq. \eref{rhoPhI} is non-negative definite, namely $\gamma > \gamma_c$, 
the previous solution ceases to be valid, since the inequality $%
\rho (\varphi) \ge 0$ is broken at certain values of $\varphi$. Therefore we
have to come back to the saddle point equation \eref{SPErho} and find a new
solution dropping the assumption $\mathrm{supp} \rho = [- \pi, \pi]$.
Instead of such condition, we assume that the eigenvalue density $\rho$ is
supported on $\ell$ disconnected arcs of the unit circle: 
\begin{equation}
\mathrm{supp} \rho = \bigcup\limits _{k=0} ^{\ell-1} \left[ \phi_k ^{-} ,
\phi_k ^{+} \right] .
\end{equation}
Solutions of that type are called $\ell$-cut solution, the arcs $\left[
\phi_k ^{-} , \phi_k ^{+} \right]$ on which $\rho$ is supported are
called ``cuts'', while the complementary arcs $] \phi_k ^{+} , \phi_{k+1}
^{-} [$ on which the $\rho$ identically vanishes are called ``gaps''. Here
we briefly review the general strategy to solve the singular integral
equation \eref{SPErho} when the support of the eigenvalue density does not
cover the whole circle, mainly following \cite[%
Appendix B]{MinwallaWadia}.

The first step is to use the standard change of variables $z=e^{\mathrm{i}
\varphi}$, $u= e^{\mathrm{i} \vartheta}$ and name the boundaries of the cuts 
$A_k = e^{\mathrm{i} \phi_k ^{-}}, B_k = e^{\mathrm{i} \phi_k ^{+}}$, $%
k=0, \dots, \ell-1$, so that the saddle point equation \eref{SPErho} is
rewritten as: 
\begin{equation}
- \gamma V^{\prime} (z, z^{-1}) = \mathrm{i}\sum_{k=0} ^{\ell -1} \mathrm{P}
\int_{A_k} ^{B_k} \frac{ \mathrm{d} u }{ \mathrm{i} u } \psi (u) \frac{ z +u%
}{z -u} ,  \label{SPEpsi}
\end{equation}
where $\psi (e^{\mathrm{i} u}) \equiv \rho (\vartheta)$ and $V^{\prime} (z,
z^{-1})$ is the derivative of the potential \eref{Vgen} written in terms of
the new variable. We now introduce the function of complex variable $z \in 
\mathbb{C}$: 
\begin{equation}
\Psi (z) = \sum_{k=0} ^{\ell -1} \int_{A_k} ^{B_k} \frac{ \mathrm{d} u }{ 
\mathrm{i} u } \psi (u) \frac{ z +u}{z -u} ,
\end{equation}
with integrals taken along the cuts on the unit circle in $\mathbb{C}$. When 
$z$ belongs to the unit circle, that is $e^{\mathrm{i} \varphi}$ for some $%
\varphi \in \mathrm{supp} \rho$, from the very definition we have 
\begin{equation}
\lim_{ z \downarrow e^{\mathrm{i} \varphi} } \Psi (z) - \lim_{z \uparrow e^{%
\mathrm{i} \varphi} } \Psi (z) \equiv \Psi_{+} (e^{\mathrm{i} \varphi}) -
\Psi_{-} (e^{\mathrm{i} \varphi}) = 4 \pi \psi (e^{\mathrm{i} \varphi}) ,
\end{equation}
where $\Psi_{+} (z)$ (respectively $\Psi_{-} (z)$) denotes the limit from
outside (respectively inside) the unit circle. We also define the auxiliary complex function 
\begin{equation}
h(z) = \sqrt{ \prod_{k=0} ^{\ell -1} ( A_k - z)( B_k - z) } ,
\end{equation}
which has branch cuts exactly along the arcs on which $\rho ( \varphi)$ is
supported (i.e., the cuts). Consider also the function $\Phi (z)$ such that 
\begin{equation}
\Psi (z) = h(z) \Phi (z), \quad z \in \mathbb{C} .
\end{equation}
Standard calculations based on the application of the residue theorem and
simple contour manipulations provide the identity 
\begin{equation}
\Phi (z) = \frac{1}{2 \pi} \oint _{C} \frac{ - \gamma V^{\prime} (u, u^{-1}) 
}{h(u) (u-z)} \mathrm{d} u ,  \label{Phiintegral}
\end{equation}
where the integration contour $C$ encloses the $\ell$ cuts but not the point $%
z$, and we have taken into account the relation between $\Psi (z) $ and $%
\Phi (z)$ and the fact that $\Psi (z)$ satisfies the saddle point equation %
\eref{SPEpsi} along the $\ell$ cuts.

At this point we have all the ingredients to obtain the function $\psi (e^{%
\mathrm{i} \varphi})= \rho (\varphi)$, but more information about the
potential $V(z,z^{-1})$ is required in order to fix the number $\ell$ of
cuts and to evaluate the integral \eref{Phiintegral}. Under the hypothesis
that $V^{\prime} (z, z^{-1})$ has poles but not branch cuts in the complex
plane, the integral expression \eref{Phiintegral} can be manipulated \cite%
{MinwallaWadia} to obtain: 
\begin{equation}
\Phi (z) = \mathcal{I}_1 (z) + \mathcal{I}_2 (z) + \mathcal{I}_3 (z) ,
\label{PhiSum3}
\end{equation}
where $\mathcal{I}_1 (z) $ computes the residue in $z$, $\mathcal{I}_2 (z) $
computes the residues at the poles of $V^{\prime}$ and $\mathcal{I}_3 (z) $
is the residual contour integration along a very large circle, used to avoid
the branch cuts of $h(u)$. Explicitly: 
\begin{eqnarray}
\mathcal{I}_1 (z) & = - \gamma \frac{ V^{\prime} (z,z^{-1}) }{\mathrm{i} h
(z) } , \\
\mathcal{I}_2 (z) & = - \gamma \sum_{z_r} \mathrm{Res}_{u=z_r} \frac{
V^{\prime} (u,u^{-1}) }{\mathrm{i} h (u) (z-u) } , \\
\mathcal{I}_3 (z) & = \lim_{R\to \infty} \frac{1}{2 \pi} \oint _{C_R} \frac{
- \gamma V^{\prime} (u, u^{-1}) }{h(u) (u-z)} \mathrm{d} u ,
\end{eqnarray}
where the sum in $\mathcal{I}_2$ runs over the poles of $V^{\prime} (z,
z^{-1})$ and the contour $C_R$ in the definition of $\mathcal{I}_3$ is a
large circle of radius $R$. The number $\ell$ of cuts must be consistent
with $\mathcal{I}_3$ to be finite or vanish. Notice that, if $V (z, z^{-1})$
is smooth in the unit circle, $\mathcal{I}_1$ is irrelevant for the
calculation of $\psi (z)$, since in that case $h(z) \mathcal{I}_1 (z)$
yields no jump.

\subsection{Specializations at large $\protect\gamma$}

We now provide explicit formulae in the cases relevant for the forthcoming sections.

\begin{figure}[htb]
\centering
\includegraphics[width=0.9\textwidth]{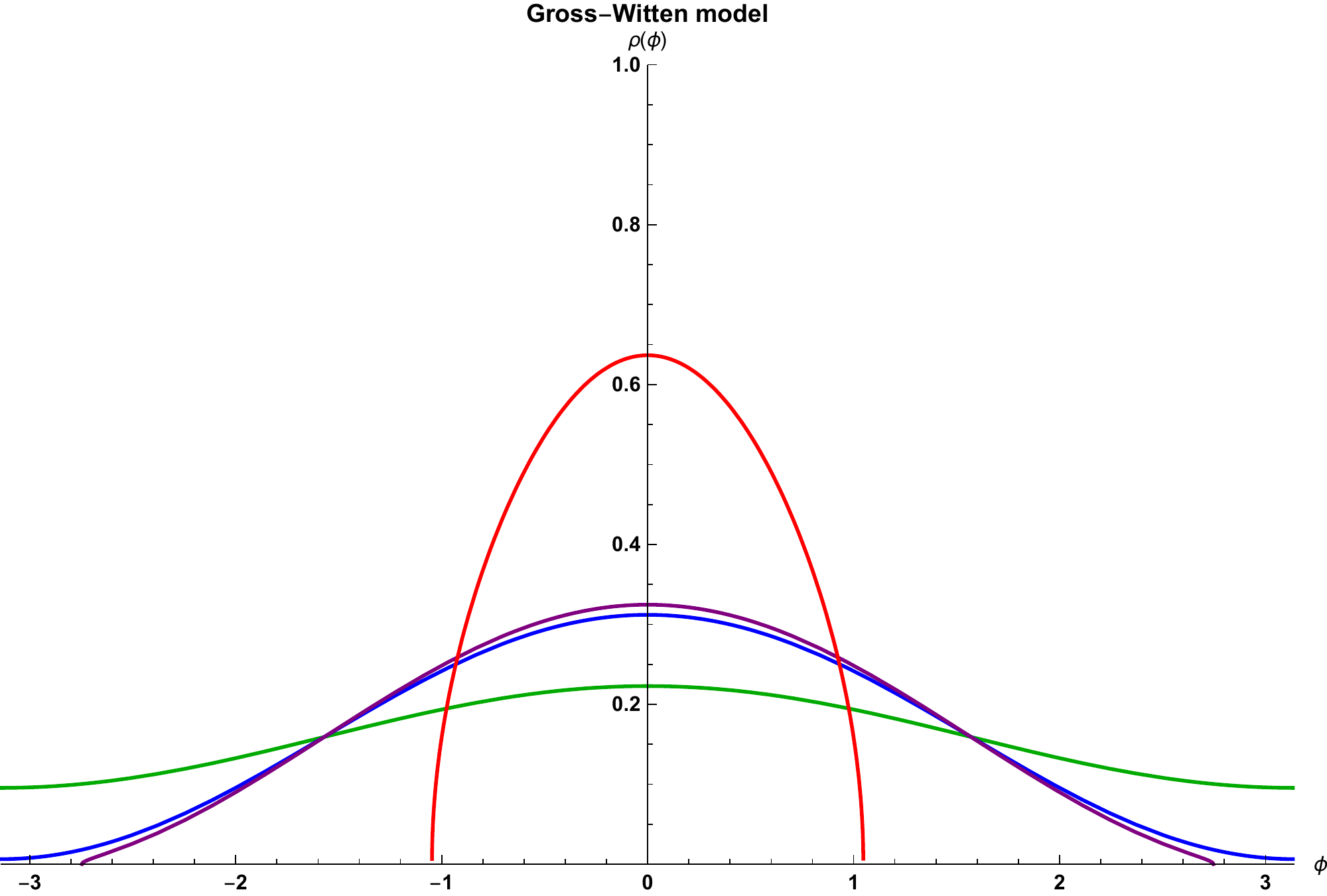}
\caption{Eigenvalue density in the Gross-Witten model, for different values of the control parameter $\gamma=$ $0.2$ (green), $0.48$ (blue), $0.52$ (purple), $2$ (red). The critical value is $\gamma_c=0.5$.}
\label{fig:rhogw2d}
\end{figure}

For a spin chain with interaction only to the closest neighbour, $g_1=1$ and
all other coefficients set to zero, we get the Gross-Witten-Wadia model,
with 
\begin{equation}
V^{\prime} (z,z^{-1}) = \mathrm{i} (z-z^{-1})
\end{equation}
Looking for a symmetric one-cut solution, supported in $[- \phi_0, \phi_0]$,
we can directly obtain $\Phi (z)$ from the formula \eref{PhiSum3}. Since $%
V^{\prime}$ has only a pole in $z=0$ and no branch cuts, thus $\mathcal{I}_1 
$ is irrelevant for the evaluation of $\psi (z)$. The other two
contributions are: 
\begin{eqnarray}
\mathcal{I}_2 (z) & = - \frac{ \gamma }{ h(0) z } = \gamma z^{-1} , \\
\mathcal{I}_3 (z) & = \lim_{R\to \infty} \oint _{C_R} \frac{ - \ii \gamma \dd u }{2 \pi u } = \gamma , 
\end{eqnarray}
where for the second equality in the evaluation of $\mathcal{I}_2 (z)$ we
used the fact that $h(z)$ approaches the real axes with positive sign if $%
|z|>1$ but with negative sign if $|z|<1$, so in particular $h(0)=-1$. We
therefore obtain 
\begin{equation}
\Psi_{+} (e^{\mathrm{i} \varphi}) - \Psi_{-} (e^{\mathrm{i} \varphi}) =
\gamma (1+e^{-\mathrm{i} \varphi}) \left( h_{+} (e^{\mathrm{i} \varphi}) -
h_{-} (e^{- \mathrm{i} \varphi}) \right) = 2 \gamma (1+e^{-\mathrm{i}
\varphi}) h_{+} (e^{\mathrm{i} \varphi}) .
\end{equation}
Therefore we obtain the Gross-Witten eigenvalue density at strong coupling 
\cite{GW}: 
\begin{equation}
\rho_{\mathrm{GW}} (\varphi) = \frac{ \gamma }{\pi} \cos \frac{ \varphi}{2} \sqrt{ \left(
\sin \frac{ \phi_0 }{2} \right)^2 - \left( \sin \frac{ \varphi }{2}
\right)^2 } ,
\end{equation}
with 
\begin{equation}
\left( \sin \frac{ \phi_0 }{2} \right)^2 = \frac{1}{2 \gamma}
\end{equation}
fixed by normalization. The Gross-Witten eigenvalue density at different values of $\gamma$ is plotted in \fref{fig:rhogw2d}.

\begin{figure}[htb]
\centering
\includegraphics[width=0.9\textwidth]{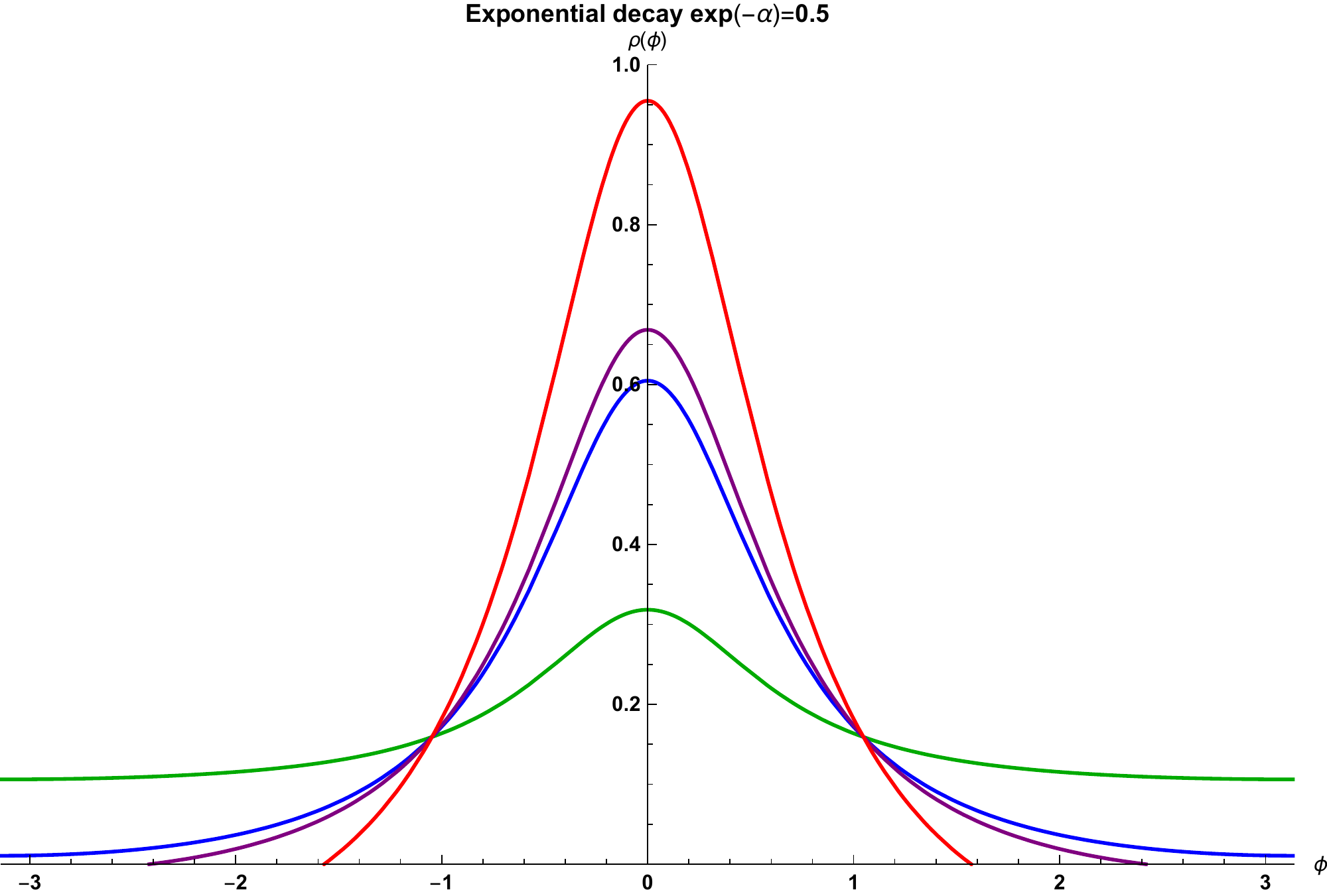}
\caption{Eigenvalue density for exponentially decaying interaction, at $e^{- \alpha} = 0.5$, for different values of the control parameter $\gamma=$ $0.5$ (green), $1.4$ (blue), $1.6$ (purple), $2.5$ (red). The critical value for this choice of $\alpha$ is $\gamma_c = 1.5$.}
\label{fig:rhoBaik2d}
\end{figure}

\begin{figure}[htb]
\centering
\includegraphics[width=0.7\textwidth]{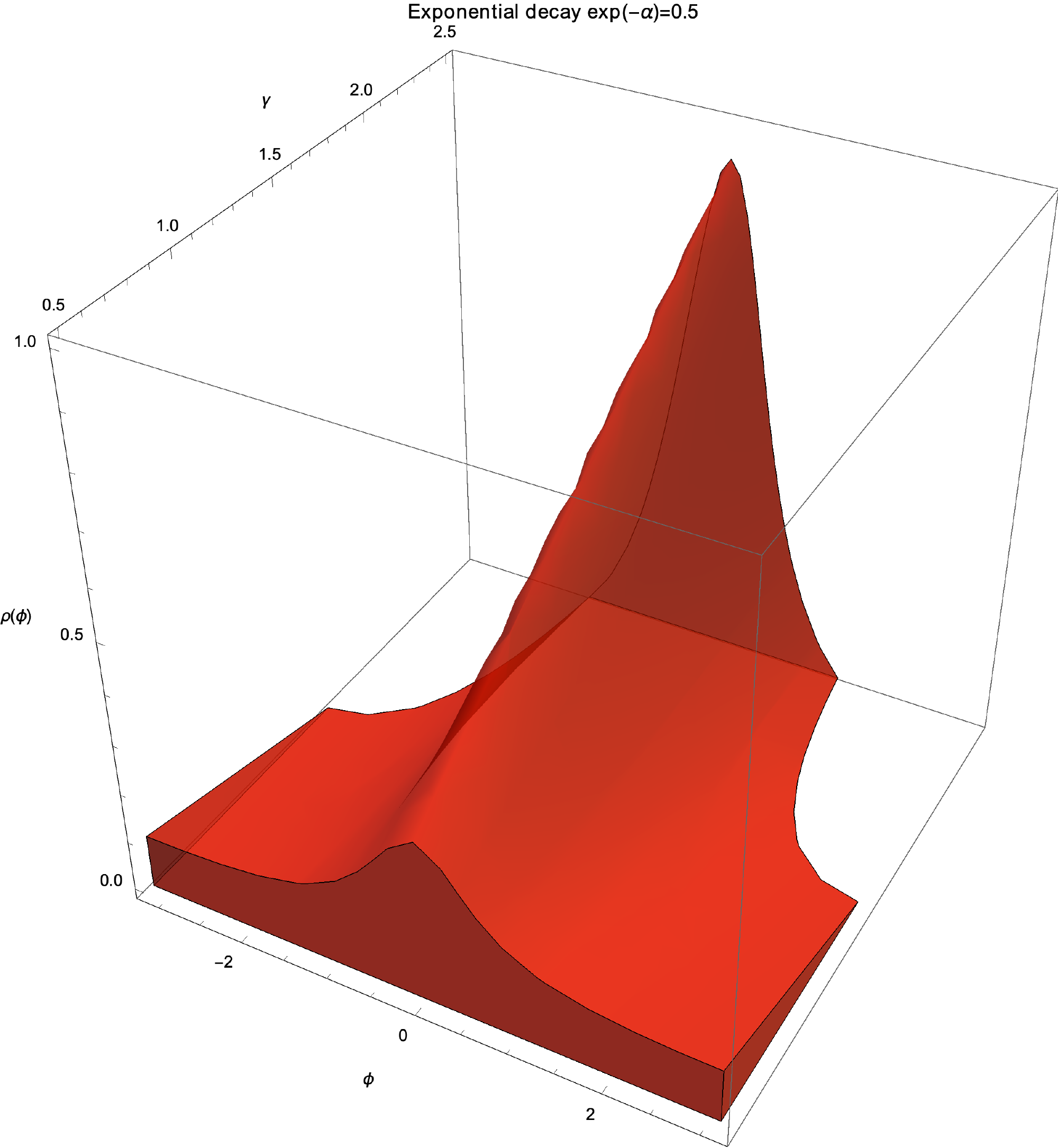}
\caption{Eigenvalue density for exponentially decaying interaction, at $e^{- \alpha} = 0.5$, as a
function of $\protect\gamma$. The phase
transition is signalled by the shrinkage of the support.}
\label{fig:rhoBaik3d}
\end{figure}

The next example corresponds to a spin chain with exponentially decaying
interaction, as in \eref{VBaik}. In this case we have: 
\begin{equation}
- \gamma V^{\prime} (z, z^{-1}) = - \mathrm{i} \gamma e^{- \alpha} \left[ 
\frac{z}{1- e^{- \alpha} z } - \frac{z^{-1} }{1- e^{- \alpha} z^{-1} } %
\right]
\end{equation}
Again $V^{\prime} (z, z^{-1})$ has poles but no branch cuts, so we can apply
formula \eref{PhiSum3}. Again $\mathcal{I}_1$ is irrelevant for the
calculation of $\rho (\varphi)$, while $\mathcal{I}_3 =0$. Thus the unique
contribution comes from the poles of $V^{\prime}$ at $e^{\mp \alpha}$: 
\begin{eqnarray}
\fl \mathcal{I}_2 (z) = - \gamma \left[ \frac{e^{-\alpha}}{h(e^{-\alpha})(z-e^{-%
\alpha})} - \frac{1}{h(e^{\alpha})(1-e^{-\alpha} z)} \right] = \frac{ \gamma 
}{h (e^{- \alpha})} \left[ \frac{e^{- \alpha} }{ e^{- \alpha} - z} - \frac{1%
}{e^{\alpha} - z} \right] ,
\end{eqnarray}
where for the second equality we used: 
\begin{equation}
h(t^{-1}) = \sqrt{ 1 + t^{-2} - 2t^{-1} \cos \phi_0 } = - t^{-1} h (t) .
\end{equation}
We then get: 
\begin{eqnarray}
\rho (\varphi) & = \frac{1}{4 \pi} \mathcal{I}_2 (e^{\ii \varphi}) ( h_+ (e^{\ii \varphi}) - h_{-} (e^{\ii \varphi} ) ) = \frac{h_+ (e^{\ii \varphi} ) }{2 \pi} \mathcal{I}_2 (e^{\ii \varphi}) \nonumber \\ 
	& = \frac{ \gamma e^{\ii \varphi /2} }{\pi h (e^{- \alpha}) } \sqrt{ \left( \sin \frac{ \phi_0 }{2} \right)^2 - \left( \sin \frac{ \varphi }{2} \right)^2 } \left[ \frac{e^{- \alpha} }{ e^{- \alpha} - e^{\ii \varphi} } - \frac{1}{e^{\alpha} - e^{\ii \varphi}} \right] ,
\label{rhoBaikIIa}
\end{eqnarray}
with boundary of the support fixed by normalization: 
\begin{equation}
\left( \sin \frac{ \phi_0 }{2} \right)^2 = \frac{ (2 \gamma -1) (1 - e^{-
\alpha} )^2 }{4 e^{- \alpha} (\gamma-1)^2 } .
\end{equation}
As a consistency check, we see that when $\gamma \to \gamma_c$ from above, $%
\phi_0 \to \pi$, meaning that the solution at small $\gamma$ is approached
by removing the hard wall at $\phi_0$. Inserting this expression into %
\eref{rhoBaikIIa} allows to write $\gamma / h (e^{- \alpha})$ in a simpler
form, and we finally get: 
\begin{equation}
\rho (\varphi) = \frac{ ( \gamma -1)(1+ e^{\mathrm{i} \varphi}) h_{+} (e^{%
\mathrm{i} \varphi}) }{ 2 \pi (e^{\mathrm{i} \varphi} - e^{- \alpha})(e^{%
\mathrm{i} \varphi} - e^{\alpha}) } .
\end{equation}
This provides an alternative derivation of the result of Baik also at large $%
\gamma$ \cite{JB}. We plot the eigenvalue density for different values of $\gamma$ in \fref{fig:rhoBaik2d}, and also show how it changes when $\gamma$ is continuously varied, in \fref{fig:rhoBaik3d}.

We can thus use this expression to obtain the large $N$ limit of the free
energy of the system in the strong coupling phase, $\gamma>\gamma_c$. We replace $e^{- \alpha} \to t$, where $t$ is now a variable, and take the derivative
of $\mathcal{F}$ with respect to $t$. This gives: 
\begin{eqnarray}
\fl \frac{ \dd \mathcal{F} }{ \dd t} (\gamma > \gamma_c) & =
\gamma \int_{- \phi_0} ^{\phi_0} \mathrm{d} \varphi \rho (\varphi) \left[
\frac{ e^{\ii \varphi} }{1- t e^{\ii \varphi} } + \frac{ e^{- \ii \varphi}
}{1- t e^{- \ii \varphi} } \right] \nonumber \\ 
& = 2 \gamma (\gamma -1) t \int_{-
\phi_0} ^{\phi_0} \frac{ \dd \varphi }{\pi} \frac{ \cos \frac{ \varphi}{2}
\sqrt{ \left( \sin \frac{ \phi_0 }{2} \right)^2 - \left( \sin \frac{ \varphi
}{2} \right)^2 } }{ (1-t)^2 + 4t \left( \sin \frac{ \varphi}{2} \right)^2 }
\left[ \frac{ 2 \cos \varphi - 2 t }{ 1 +t^2 -2t \cos \varphi } \right] \nonumber \\ 
& = \frac{ 8 \gamma (\gamma-1) t }{\pi} \int_{-x_0} ^{x_0} \mathrm{d} x \frac{
\sqrt{ x_0^2 - x^2 } (1-t -2x^2) }{ [ (1-t)^2 +4 t x^2 ]^2 } \nonumber \\
& = \frac{ \gamma (\gamma -1)}{\pi t } \int_{-1} ^{1} \mathrm{d} y \frac{ \sqrt{1-y^2}
\left[ \frac{ 2t (\gamma -1)^2 }{(1-t)(2\gamma-1)} -y^2\right] }{\left[
\frac{ (\gamma-1)^2 }{2 \gamma -1} + y^2 \right]^2 } \nonumber \\ 
& = \frac{ 1 + t - 4 \gamma t }{2 t (1-t) } , 
\end{eqnarray}
where we first changed variables $x = \sin \frac{\varphi}{2} $, with $x_0 =
\sin \frac{ \phi_0}{2}$, and then scaled again the variables $y= x/x_0$ and
used the explicit form of $x_0 ^2$ to simplify the expression. Integrating
this up to $t=e^{- \alpha}$ we get: 
\begin{equation}
\mathcal{F} ( \gamma > \gamma_c) =  \gamma h + (2 \gamma -1) \log (1- e^{- \alpha}) -  \frac{\alpha}{2} + \mathcal{C} (\gamma) ,
\end{equation}
where $\mathcal{C} (\gamma) $ is some $\alpha$-independent constant.

The last case we solve explicitly is the case of polylogarithmic potential %
\eref{Vpolylog}, hence 
\begin{equation}
V^{\prime} (z, z^{-1}) = \mathrm{Li}_p (e^{-\alpha} z) - 
\mathrm{Li}_p (e^{-\alpha} z^{-1} ) .
\end{equation}
The case $p=0$ reduces to the case above, and for all $p \in \mathbb{Z}_{
\le 0}$ the derivation goes through in the exactly same way, the unique
difference being that the residue in $e^{\alpha}$ has positive (respectively
negative) sign when $p$ is even (resp. odd). Therefore the eigenvalue
density $\rho (\varphi)$ for all $p \le 0$ is given by the formula %
\eref{rhoBaikIIa}, up to a factor $(-1)^p$ in the last summand in square
bracket.

The case $p \in \mathbb{Z}_{>0}$ however is more involved, since in that
case the polylogarithm has a branch cut but no poles. The eigenvalue density
in that case, for $\alpha=0$, has been obtained in \cite{AmadoSundborg},
although in a rather different context. Turning on the exponential decay $%
\alpha \ge 0$ their procedure works identically, leading to: 
\begin{eqnarray}
\fl \rho ( \varphi) = \frac{ \gamma }{\pi} \left( \frac{ \gamma }{\gamma_c}
\right)^{p-1} \left[ \mathrm{Li}_p (e^{- \alpha} e^{ \mathrm{i} \varphi}) + 
\mathrm{Li}_p (e^{- \alpha} e^{ - \mathrm{i} \varphi}) - \mathrm{Li}_p (e^{-
\alpha} e^{ \mathrm{i} \phi_0}) - \mathrm{Li}_p (e^{- \alpha} e^{ - \mathrm{i%
} \phi_0}) \right] ,
\end{eqnarray}
with $\phi_0$ implicitly determined by normalization condition. We can use
this expression to evaluate the derivative of the free energy with respect
to the variable $t=e^{- \alpha}$: 
\begin{eqnarray}
\fl \frac{ \dd \mathcal{F} }{ \dd t} (\gamma > \gamma_c) & = 2
\gamma \int_{- \phi_0} ^{\phi_0} \mathrm{d} \varphi \rho (\varphi)
\sum_{m=1} ^{\infty} \frac{ t^m }{m^p} \left[ \cos (m \varphi) - \cos ( m
\phi_0) \right] \nonumber \\ 
\fl & = \frac{ 4 \gamma^{p+1} }{\pi \gamma_c ^{p-1} }
\sum_{m=1} ^{\infty} \sum_{n=1} ^{\infty} \frac{ t^{m+n} }{(mn)^p} \int_{-
\phi_0} ^{\phi_0} \mathrm{d} \varphi \cos (n \varphi) \left[ \cos (m
\varphi) - \cos ( m \phi_0) \right] \nonumber \\ 
\fl & = \frac{ 4 \gamma^{p+1} }{\pi
\gamma_c ^{p-1} } \sum_{m=1} ^{\infty} \sum_{n=1} ^{\infty} \frac{ t^{m+n}
}{(mn)^p} \nonumber \\
\fl & \times \left[ \frac{ \sin ( \phi_0 (m+n))}{n+m} + \frac{ \sin ( \phi_0
(m-n))}{n-m} - \frac{ 2 \cos ( m \phi_0) \sin (n \phi_0) }{n} \right] .
\label{dFdtPolylogII}
\end{eqnarray}

\section{XX model with asymmetric hopping}
\label{sec:GWasym}
We consider a generalization of the XX model case discussed in \cite{DM}, which corresponds to the Gross-Witten matrix model, whose potential is\footnote{We follow the matrix model convention, which may differ by a factor 2 from the literature discussing the XX spin chain.} $V(\varphi)= h+ 2 \cos (\varphi )$ \cite{DM}, and consider the matrix model potential
\begin{equation}
V(\varphi )=h+2\left( \cos (\varphi )-\ii v \sin (\varphi )\right) .
\label{Viv}
\end{equation}
More specifically, this modifies the hopping term in \eref{XX-o} in the following way 
\begin{equation}
\hat{H}_{\mathrm{hop}}=-\frac{1}{2}\sum_{i =0 }^{\infty}  \left( (1- v)\sigma _{i}^{-}  \sigma
_{i+1}^{+}+(1+ v)\sigma _{i}^{-}  \sigma _{i-1}^{+}\right) ,
\label{HhopGWv}
\end{equation} 
where the parameter $v$ measures the asymmetry between interaction on the right and on the left. We remark that, in the definition of the Hamiltonian $\hat{H}_{\mathrm{hop}}$ in \eref{HhopGWv}, we use the Pauli matrices $\sigma^{\pm}$, and therefore the model is different from the XY spin chain, whose Hamiltonian has similar form but the matrices $\sigma^{x}, \sigma^{y}$ are used. 

In \cite{Godreche,GodLuck} for example, this model was considered for the particular case of a single spin flip (Glauber
dynamics). Already at the one spin flip level, two different behaviours were observed 
depending on whether $0\leq v\leq 1$ or $v>1$. This is due to the representation of correlation functions as modified Bessel functions of first kind, $I_n (z)$, whose argument is real for $0 \le v \le 1$, but it becomes imaginary  when $v>1$.
Through the relation $\ii ^{n} I_n (- \ii x) =  J_n (x)$ \cite{Bessel}, one can write a modified Bessel function with pure imaginary argument as an ordinary Bessel function of real argument. The modified Bessel functions $I_n (x)$ are monotone, while the ordinary Bessel functions $J_n (x)$ are damped oscillating functions.
This effect was responsible for the presence of two regimes in \cite{GodLuck}.

On the other hand, at the matrix model level, the Gross-Witten model (which describes the amplitude \eref{eq:Zamplw} with nearest-neighbour interaction) shows a large $N$ phase transition for imaginary time $w=\beta$, while there is a unique phase for real time $w= \ii t$.
Such aspect is tightly related to the Fourier coefficients of the symbol being modified Bessel functions $I_n (2w)$, hence passing from imaginary- to real-time dynamics changes the behaviour of the partition function (see subsection \ref{app:Bessel+v}). This motivates us to investigate the case of a $N$-spin
flip process in the present formalism, for potential \eref{Viv}, to establish a relation between imaginary-time dynamics at $0\le v \le 1$ and real-time dynamics at $v>1$.

\subsection{Determinants of Bessel functions}
\label{app:Bessel+v}
\begin{figure}[hbt]
\centering
\includegraphics[width=0.45\textwidth]{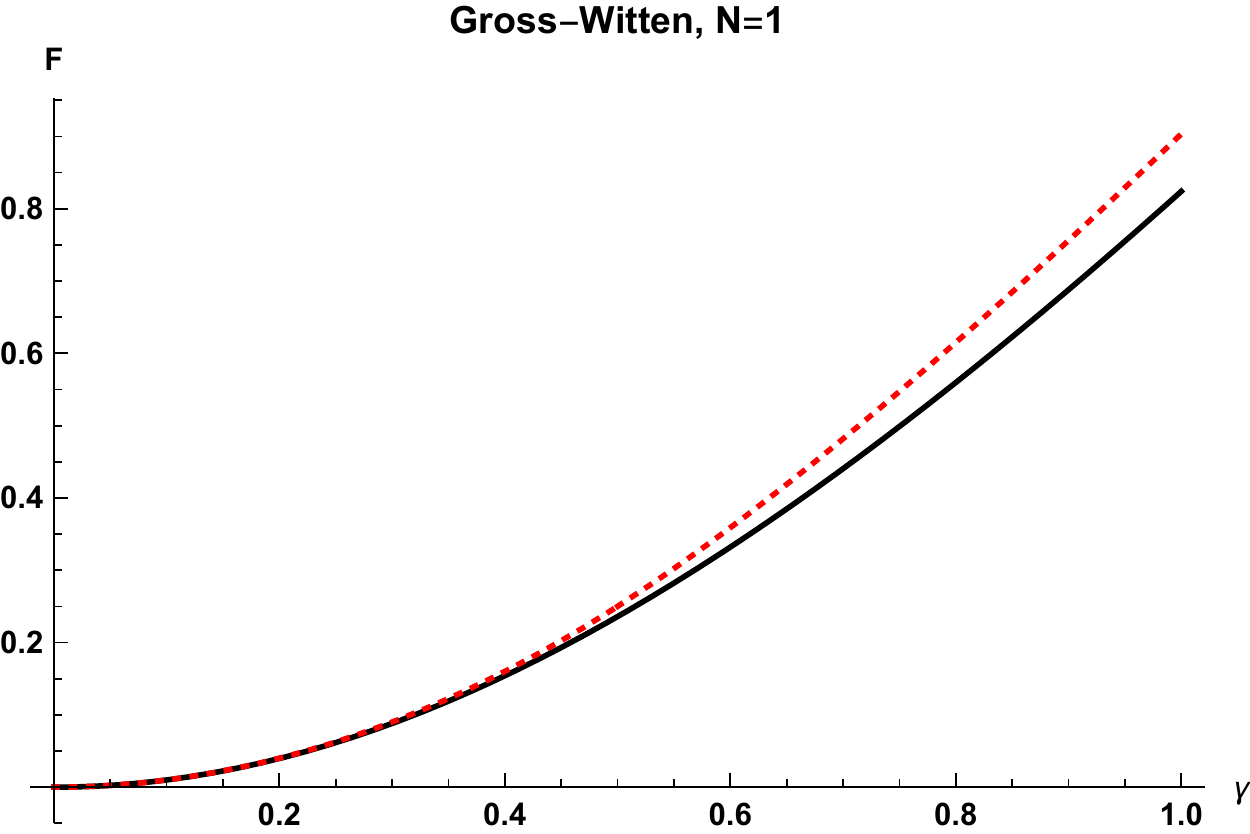}%
\hspace{0.08\textwidth}%
\includegraphics[width=0.45\textwidth]{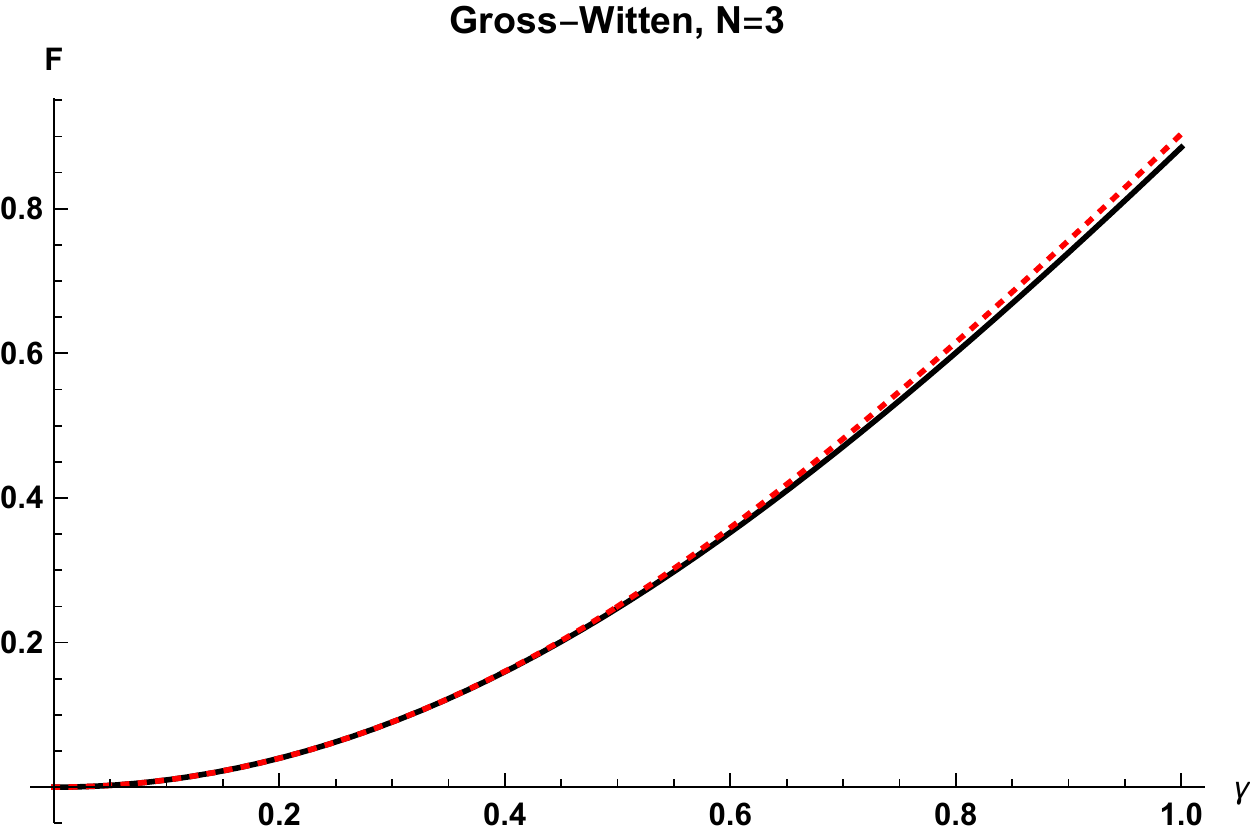}\\
\vspace{0.5cm}
\includegraphics[width=0.45\textwidth]{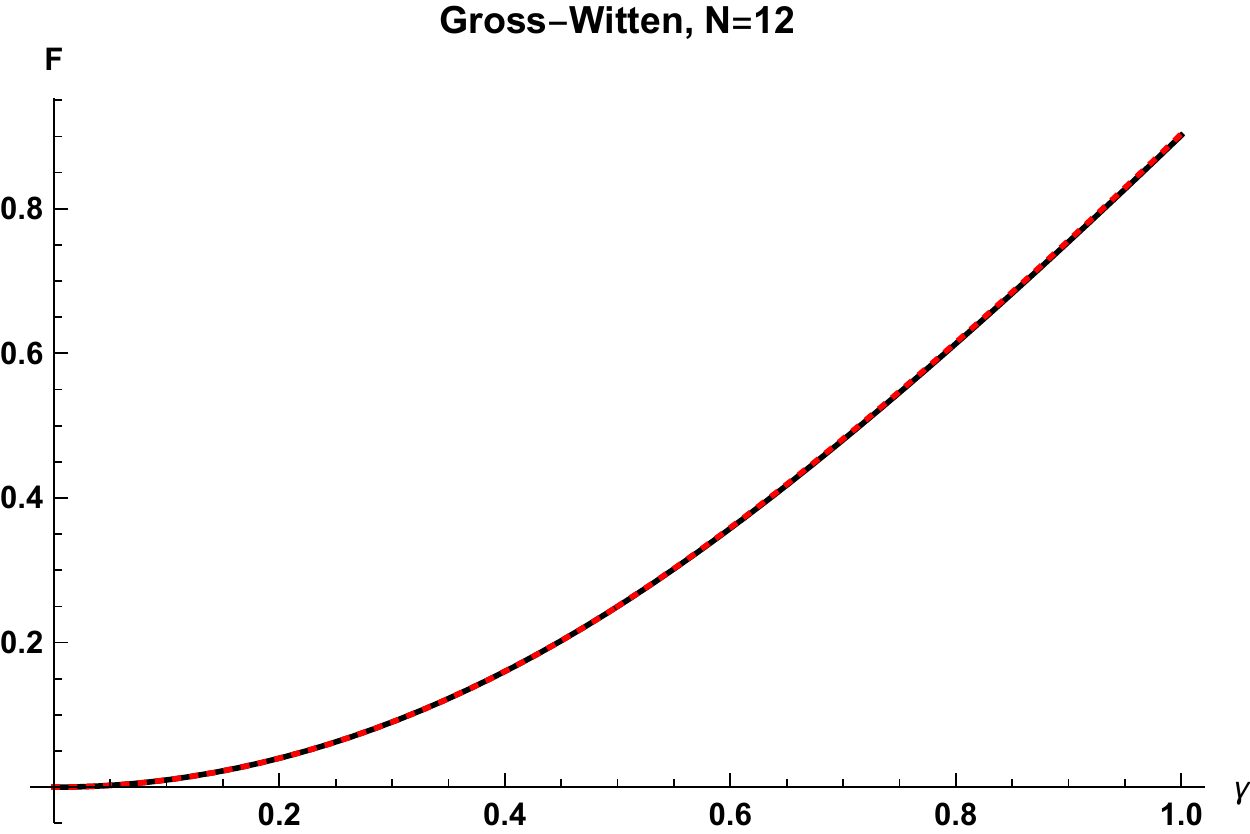}
\caption{Comparison between the matrix model free energy $\mathcal{F}_{\mathrm{GW}}$ of the Gross-Witten model computed through the determinant of modified Bessel functions of first kind \eref{ZGWdet} (black) and the theoretical large $N$ formula \eref{FGW} (red, dotted) as a function of $\gamma$, for $N=1$ (up left), $N=3$ (up right) and $N=12$ (center).}
\label{fig:convergenceGW}
\end{figure}

We briefly review the representation of the XX model and generalizations thereof in terms of a determinant of modified Bessel functions of first kind, denoted by $I_n (z)$.
The complex-time partition function $\mathcal{Z} (w)$ associated to the XX model is:
\begin{equation}
\mathcal{Z} (w) = \frac{ e^{whN}}{N!} \int_{-\pi} ^{\pi} \frac{ \dd \varphi_1}{2 \pi} \cdots \int_{-\pi} ^{\pi} \frac{ \dd \varphi_N}{2 \pi} \prod_{1 \le j < k \le N} \left\vert e^{\ii \varphi_j} - e^{\ii \varphi_k} \right\vert e^{2 w \sum_{j=1} ^N \cos (\varphi_j) } ,
\label{ZGWmat}
\end{equation}
which, for purely real parameter $w=\beta$ (imaginary time) reduces to the celebrated Gross-Witten-Wadia model \cite{GW,Wadia}. From the identity
\begin{equation}
\exp \left\{ w \left( \frac{z + z^{-1}}{2} \right) \right\} = \sum_{n \in \mathbb{Z}} I_n (w) z^{n} 
\label{BesselFourier}
\end{equation}
and the relation between unitary matrix model and Toeplitz determinant, we get:
\begin{equation}
\mathcal{Z} (w) = e^{whN} \det \left[ I_{j-k} (2w) \right]_{j,k=1} ^N
\label{ZGWdet}
\end{equation}
For imaginary-time dynamics $w=\beta$, the partition function \eref{ZGWdet} shows a third order phase transition at large $N$, with $\beta/N \equiv \gamma $ fixed \cite{GW}. In particular, the free energy $\mathcal{F} = \log \mathcal{Z} (\beta) / N^2$ is given by:
\begin{equation}
\mathcal{F}_{\mathrm{GW}} (\gamma) = \cases{  \gamma^2 ,  & $ \gamma \le \frac{1}{2} ,$ \\  2 \gamma - \frac{1}{2} \log 2 \gamma - \frac{3}{4} ,  &$ \gamma > \frac{1}{2} . $ \\ }  
\label{FGW}
\end{equation}
Such result, obtained using saddle point techniques, is a particular case studied above in section \ref{app:saddleptsmethod}. The same result may also be proved through a more powerful procedure, namely solving a Riemann-Hilbert problem \cite{BDJ}. The convergence of the logarithm of the determinant \eref{ZGWdet} to the theoretical large $N$ formula \eref{FGW} as $N$ increases is showed in \fref{fig:convergenceGW}.

\begin{figure}[hbt]
\centering
\includegraphics[width=0.47\textwidth]{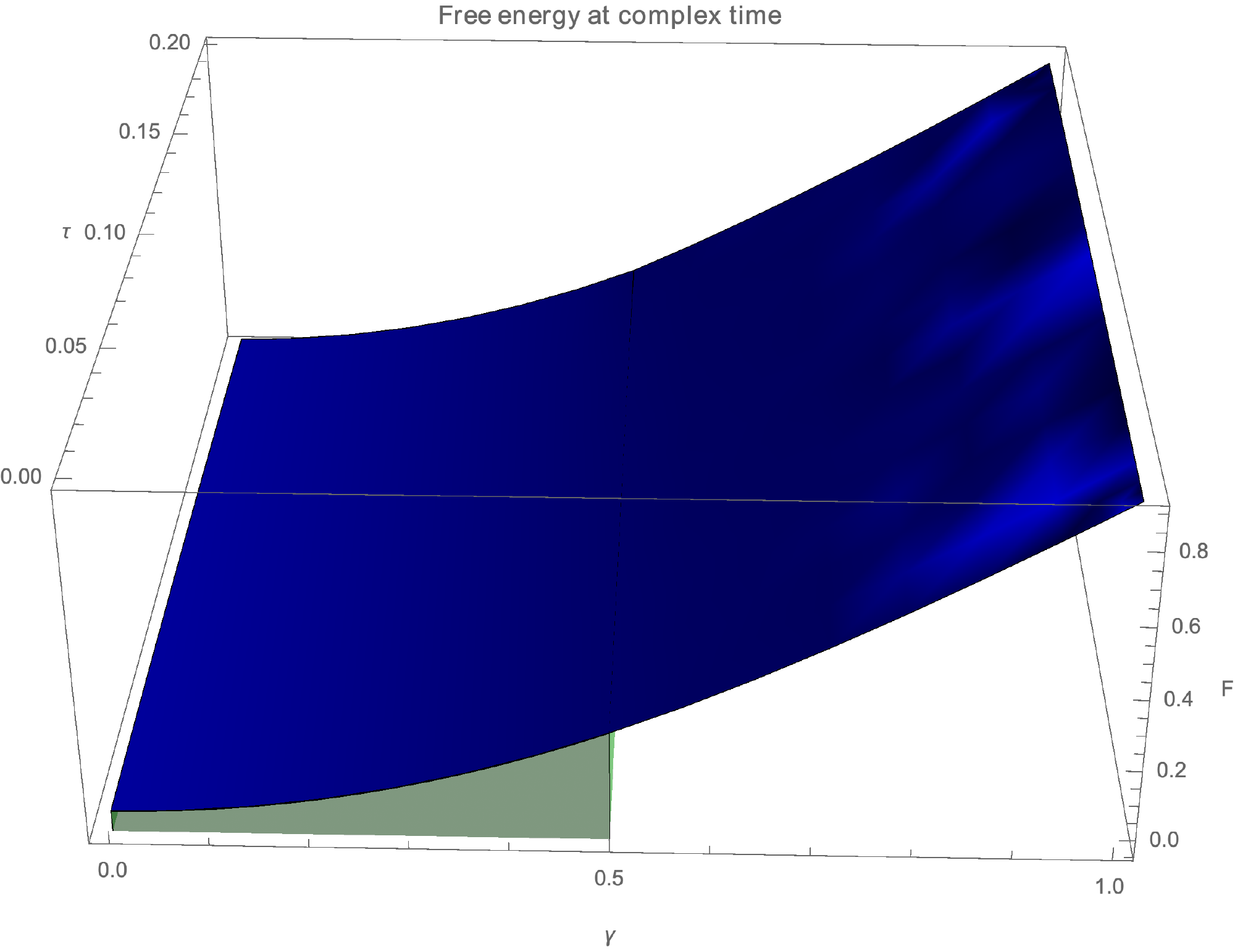}\hspace{0.04\textwidth}
\includegraphics[width=0.47\textwidth]{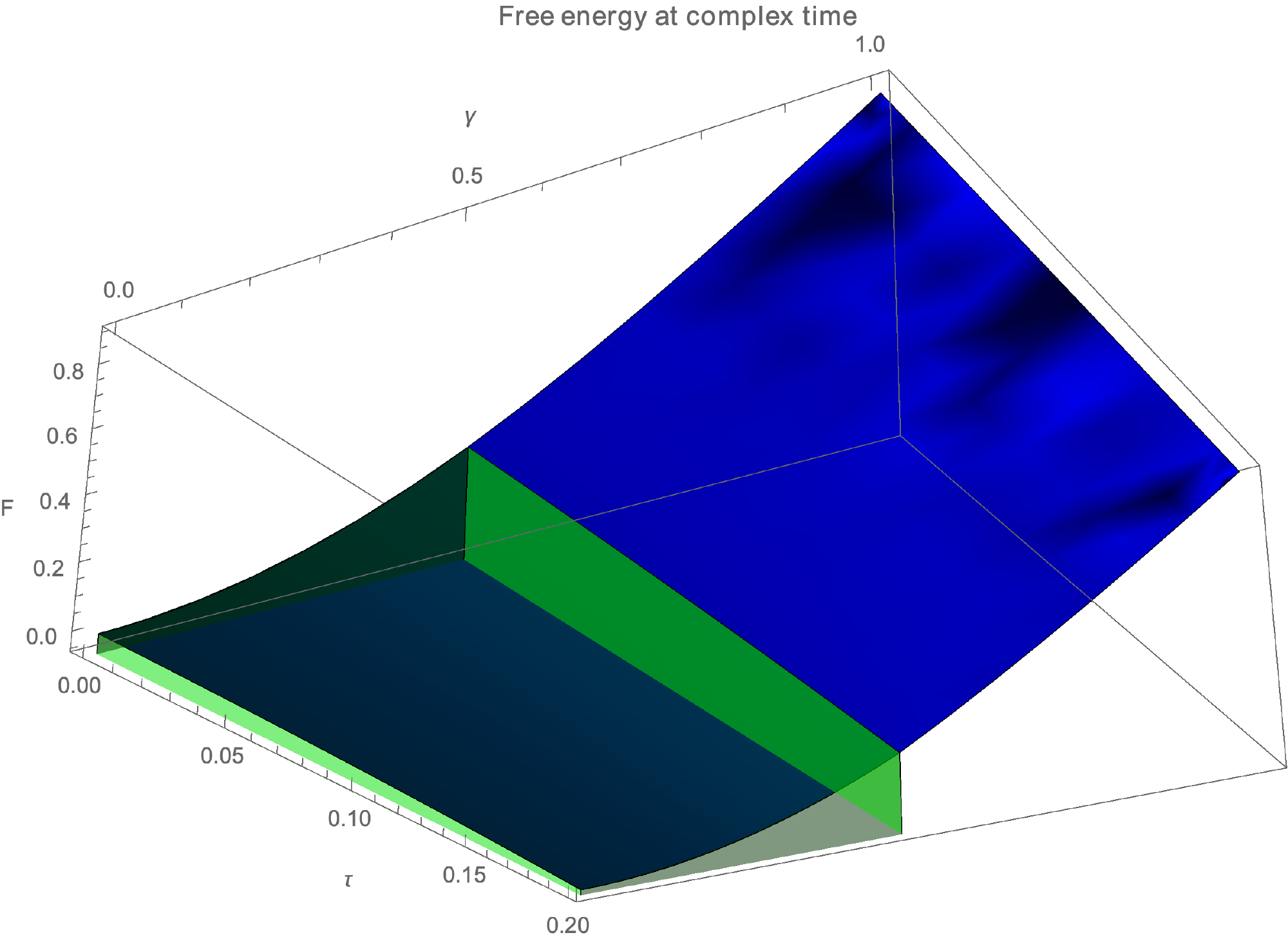}
\caption{Logarithm of the Loschmidt amplitude \eref{eq:Zamplw} for XX spin chain at complex time. Two views of the same plot, from above (left) and below (right). The blue surface shows the direct calculation of $\mathcal{F}$ from the determinantal representation at $N=10$, and the solid green region is the volume contained in the theoretical prediction, with unscaled $t$.}
\label{fig:FcomplexGw}
\end{figure}

For imaginary argument $w=\ii t$, however, the determinant \eref{ZGWdet} has a different behaviour, following from the identity \cite{Bessel}
\begin{equation}
I_n (2 \ii t) = \ii ^{- n} J_n (-2 t) ,
\end{equation}
where $J_n $ is the ordinary Bessel function of first kind. In case of complex argument $w=\beta + \ii t$, the free energy $\mathcal{F}$ shows a Gross-Witten behaviour for each slice at fixed $t$, while it is smooth for large even $N$ along its $\beta =0$ section. At large $N$ with $\beta/N$ fixed but unscaled $t$, $\mathcal{F}$ can be evaluated from the extension of Szeg\H{o} theorem described in the \hyperref[propszego]{Proposition}. This is plotted in \fref{fig:FcomplexGw}. Therefore, the presence of a phase transition at large $N$ relies on the real part of the argument being nonzero.
In the formalism of saddle point approximation at large $N$, described in detail in section \ref{app:saddleptsmethod}, this can be restated as follows: only the real part of the action admits nontrivial saddle points. As a consequence, the change of solution of the saddle point equation after a critical point in parameter space is necessarily controlled by $\gamma$.

\subsection{Determinants in the case of asymmetric hopping}
We now focus on a generalization of the model above, introducing asymmetry between the interaction on the left and on the right, see \eref{HhopGWv}. Interactions of that type have been studied in \cite{Godreche,GodLuck}, for the case of a single spin flip in the Ising chain. This corresponds to the potential described in \eref{Viv}, which leads to the matrix model for the amplitude \eref{eq:Zamplw} :
\begin{eqnarray}
\fl \mathcal{Z} (w) = \frac{ e^{whN}}{N!} \int_{-\pi} ^{\pi} \frac{ \dd \varphi_1}{2 \pi} \cdots \int_{-\pi} ^{\pi} \frac{ \dd \varphi_N}{2 \pi} \prod_{1 \le j < k \le N} \left\vert e^{\ii \varphi_j} - e^{\ii \varphi_k} \right\vert e^{2 w \sum_{j=1} ^N \left[\cos (\varphi_j) - \ii v \sin (\varphi_j) \right] } ,
\end{eqnarray}
where the asymmetry is controlled by the parameter $v$, and at $v=0$ we recover the matrix model \eref{ZGWmat}. For $0\le v \le 1$, we can rewrite
\begin{equation}
2 \left( \cos (\varphi) - \ii v \sin (\varphi) \right) = \sqrt{1-v^2} \left( e^{ \ii \varphi - \phi_v } + e^{ -(\ii \varphi - \phi_v) } \right),
\end{equation}
where the angle $\phi_v$ is defined through
\begin{equation}
\cosh (\phi _{v})=\frac{1}{\sqrt{1-v^{2}}},\quad \sinh (\phi _{v})=\frac{v}{\sqrt{1-v^{2}}}.
\end{equation}
We can then apply identity \eref{BesselFourier} to obtain the Fourier coefficients of the symbol of the matrix model, which are $e^{- n \phi_v } I_n (2 w \sqrt{1-v^2})$. From the equivalence between unitary matrix models and Toeplitz determinants, we can write down the partition function $\mathcal{Z} (w)$ as 
\begin{equation}
\mathcal{Z}(w)=e^{whN}\det_{1 \le j,k \le N} \left[ e^{(k-j)\phi _{v}}I_{j-k}\left( 2(\beta + \ii t)\sqrt{1-v^{2}}\right) \right] .
\label{ZwvI}
\end{equation}
We see that the entries of the matrix are related to those of expression \eref{ZGWdet}, with a rescaling of the argument by a factor $\sqrt{1-v^2}$ and the powers of $e^{\phi_v}$ as a prefactor, which does not affect the behaviour of the determinant, since it does not introduce oscillating behaviour.
We have seen above in section \ref{app:Bessel+v} that the scaling limit with both $\beta/N$ and $t /N$ fixed is controlled only by $\gamma = \beta/N$, and therefore the conclusions for \eref{ZGWdet} hold also for the present model, in the regime $0\le v \le 1$.

Conversely, when $v>1$, we use:
\begin{equation}
2 \left( \cos (\varphi) - \ii v \sin (\varphi) \right) = - \ii \sqrt{v^2-1} \left( e^{ \ii \varphi - \ii \frac{\pi}{2} - \hat{\phi}_v } + e^{ -(\ii \varphi - \ii \frac{\pi}{2} - \hat{\phi}_v) } \right),
\end{equation}
where the angle $\hat{\phi}_v$ is defined through
\begin{equation}
\label{defhatphiv}
\cosh (\hat{\phi} _{v})=\frac{v}{\sqrt{v^{2}-1}},\quad \sinh (\hat{\phi} _{v})=\frac{1}{\sqrt{v^{2}-1}} .
\end{equation}
We again exploit identity \eref{BesselFourier} and get:
\begin{equation}
\exp \left\{  2 w \left( \cos (\varphi) - \ii v \sin (\varphi) \right)  \right\} = \sum_{n \in \mathbb{Z}} I_n (- \ii 2 w \sqrt{v^2-1}) i^{- n} e^{-n \hat{\phi}_v} e^{\ii n \varphi}  ,
\end{equation}
so we see that, when $v>1$, the argument of the modified Bessel function acquires an extra $\ii$ factor, an we are led to:
\begin{equation}
\mathcal{Z}(w)=e^{whN}\det_{1 \le j,k \le N} \left[ (\ii e^{\hat{\phi}_v})^{(k-j)} I_{j-k}\left( 2(t - \ii \beta )\sqrt{v^{2}-1}\right) \right] , 
\label{ZwvII}
\end{equation}
with $\hat{\phi}_v$ defined in \eref{defhatphiv}. 
We have that, at $v>1$, the roles of $t$ and $\beta$ are exchanged: in this case, the phase in the multiple-scaling large $N$ limit (taken forcing $N$ to be even, for simplicity) is controlled by the parameter $t/N$. Therefore, in this regime, the phase transition is triggered by real time dynamics.

\section{Extended XX model: short and long range interactions}
\label{sec:XXextended}

In \cite{PGT}, it was showed that one can consider the generic long-range 1d
spin Hamiltonian $\hat{H}_{\mathrm{Gen}}$ introduced in \eref{Hgen-0}, and which we rewrite here for clarity:%
\begin{equation}
\hat{H}_{\mathrm{Gen}}=-\sum_{i =0 } ^{\infty} \sum\limits_{n\in \mathbb{Z}} a_{n}\left(
\sigma _{i}^{-}  \sigma _{i+n}^{+}\right) +\frac{h}{2}\sum_{i=0} ^{\infty} (\sigma
_{i}^{z}-\mathbbm{1}) .   \label{Hgen-inf}
\end{equation}%
Then, the matrix model description
of the amplitude is again given by \eref{mat} but where the weight function 
$f_{w}(e^{i\varphi })$ of the random matrix ensemble is now%
\begin{equation}
f_{w}\left( e^{\ii \varphi }\right) =f_{0}\left( e^{\ii \varphi }\right) \exp
\left( w \sum_{n\in \mathbb{Z}}a_{n}e^{\ii n\varphi }\right) ,
\end{equation}%
where we have changed the parameter $\beta \in \mathbb{R}$ for the more general $w\in \mathbb{C}$. 
Therefore the additional interactions in the Hamiltonian appear in the
potential of the matrix model as the coefficients $a_{n}$ of the $\mathrm{Tr} U^{\pm
n} $ terms, where the integer value $n$ denotes nearest neighbour interaction
for $n=1$, next-to near for $n=2$, and so on. The extension to this case was
straightforward and can not be considered rigorous for the case of
infinitely many interactions, although the decay condition on the
coefficients guarantees convergence of the integrals and hence
existence of the amplitudes. See also the discussion in \cite[Appendix D]{Echofermion}. 
The convergence condition above however is not always necessary, and we will
discuss below how results on Toeplitz determinants with a symbol with a pure
Fisher-Hartwig singularity appear.

If $a_n \ne 0$ for only finitely many $n \in \mathbb{Z}$, meaning spins interact with only finitely many neighbours, one obtains a unitary matrix model with polynomial potential. The large $N$ limit of such models has been thoroughly studied in \cite{PeriwalShevitz}, and will not consider this case here. Instead, we will focus on the case $a_n \ne 0$ for all $n \in \mathbb{Z}$.

\subsection{Exponentially decaying interactions}
\label{sec:interactionBaik}
We first focus on the Loschmidt echo for interactions in \eref{Hgen-inf} given by $a_{n}=\exp \left( -n \alpha \right) /n$ with $\alpha \geq 0$, for both
cases of a long-range chain $\alpha =0$ and a short-range one with
exponentially decaying interactions $\alpha >0$, studying their free
energies and phase transitions. The matrix model expression \eref{mat} for
these interactions and with initial state in a single-domain wall configuration 
\begin{equation}
|\psi _{0}\rangle =|\underset{N}{\underbrace{\downarrow ,\downarrow
,...,\downarrow }},\uparrow ,...\rangle ,
\end{equation}%
is
\begin{eqnarray}
\fl \mathcal{Z}(w) =\frac{\mathrm{e}^{whN}}{(2\pi )^{N}N!}\int_{[-\pi,\pi]^N} 
\prod_{1\leq j<k\leq N}\left\vert e^{i\varphi _{k}}-e^{i\varphi
_{j}}\right\vert ^{2}  \prod_{j=1}^{N}\frac{\dd \varphi_{j} }{ \left[ \left( 1- e^{-\alpha + i\varphi _{j}}\right)
\left( 1- e^{- \alpha -i\varphi _{j}}\right) \right]^{w} }.  \label{Zw}
\end{eqnarray}
From the point of view of Toeplitz determinants, the corresponding symbol
then is $\sigma (z)=\left[ \left( 1-e^{-\alpha }z\right) \left( 1-e^{-\alpha
}/z\right) \right] ^{-w},$ which has been studied in different contexts, see 
\cite{BDS} for a textbook discussion. A few properties for $\mathcal{Z}(w)$
follow: we show that for imaginary-time dynamics, where $w\rightarrow \beta
\in \mathbb{R}$ there is a third order phase transition in the double scaling limit $%
N\rightarrow \infty $, $\beta / N$ fixed. In separate work \cite%
{LST}, we explain that this phase transition is the same one that
appears in the study of domino tilings \cite{Copro}.

As happens with the case of the XX chain, the phase transition does not
occur for real-time dynamics, where $w\rightarrow \ii t\in \ii \mathbb{R}$. 
The present model has one parameter more than the Gross-Witten/XX model, corresponding to the strength of the exponentially decaying interaction, namely $\alpha$. 
The Gross-Witten picture is recovered in the limit $\alpha \to \infty$, with $\beta_{\mathrm{GW}} \equiv \beta e^{- \alpha}$ fixed.

The matrix model \eref{Zw} has a weak-coupling phase for $0\leq \gamma
\leq \frac{1+e^{-\alpha }}{2e^{-\alpha }}$ and a strong-coupling phase for $\gamma >\frac{1+e^{-\alpha }}{2e^{-\alpha }}$ \cite{JB}, and the phase transition is third order. 
The derivation of this result is a particular case of the general setting solved in section \ref{app:saddleptsmethod}.

We are interested in the large $N$
free energy in each phase, defined by $\mathcal{F}=\lim_{N\rightarrow \infty }\mathcal{Z}(\beta )/N^{2}$, which
is related to the so-called rate function $g(t)$ of the Loschmidt echo \eref{L} $\mathcal{L} (t)=\exp \left( -Ng(t)\right) $ for imaginary time $\ii t \rightarrow \beta $.
In fact, we obtain it as a special case of the following result (see \hyperref[propszego]{Proposition} in section \ref{app:saddleptsmethod}): for a general class
of symbols, free energy $\mathcal{F}$ in the weak-coupling phase coincides with the result obtained from direct
application of Szeg\H{o} strong limit theorem. In other words, for all $0\leq \gamma \leq \gamma _{c}$, the free energy can be obtained from the
unscaled limit. This result for eigenvalues on the circle is consistent with
the general phase transitions analysis of \cite{FDC} on the real line.

For the present case of exponentially decaying interaction we obtain:
\begin{equation}
    \mathcal{F} = \cases{ \gamma h +  - \gamma^2 \log (1-e^{-2 \alpha}), & $ \gamma \le \frac{1 + e^{- \alpha} }{2 e^{- \alpha}} $ , \\  \gamma h + (2 \gamma -1)\log(1- e^{- \alpha}) - \frac{\alpha}{2} + \mathcal{C} (\gamma) , & $ \gamma > \frac{1 + e^{- \alpha} }{2 e^{- \alpha} } $ , }
\label{FBaikIandII}
\end{equation}
in the weak- and strong-coupling phase respectively, where $\mathcal{C} (\gamma)$ is $\alpha$-independent.

\subsection{Long-range chain: Fisher-Hartwig asymptotics and Stokes lines}
\label{sec:interactionFH}

Systems of trapped ions have been synthesized to describe the dynamics of
transverse-field Ising models of the form~\cite{Exp}%
\begin{equation}
\hat{H} =-\sum_{0 \le m < l \le L }J_{lm}\sigma _{l}^{z}\sigma _{m}^{z}-h\sum_{l=0}^{L}\sigma
_{l}^{x}\,.
\end{equation}%
Here, $L+1$ is the total number of spins.
The coupling $J_{lm}$ is approximately of long-ranged form~\cite{Exp}%
\begin{equation}
J_{lm}\approx \frac{1}{|l-m|^{\lambda }},\quad \mathrm{for}\,\ \,|l-m|\gg 1\,,
\label{Jlmexp}
\end{equation}%
with a tunable interaction exponent $\lambda $ from $\lambda =0$ up to $%
\lambda =3$. We will particularize the setting above to the case $\alpha =0$
and hence describe a XX chain (which can be roughly seen as two copies of
the Ising model above), corresponding to a interaction exponent of $\lambda
=1$. This is then a a long-range spin chain with interactions $J_{\left\vert
l-m\right\vert }=1/\left\vert l-m\right\vert $ between spins in positions $l$
and $m$.

The Loschmidt echo then for a single-domain wall configuration, has then
the random matrix representation, in the general case of a complex time
parameter%
\begin{eqnarray}
\fl \mathcal{Z}(w) =\frac{\mathrm{e}^{whN}}{(2\pi )^{N} N!}\int\limits_{\left[
-\pi ,\pi \right] ^{N}} \dd^N  \varphi \prod_{1\leq j<k\leq N}\left\vert
e^{\ii \varphi_{j}}-e^{\ii \varphi_{k}}\right\vert ^{2}  \prod_{j=1}^{N}\left[ \left( 1-e^{\ii \varphi_{j}}\right) \left( 1-e^{- \ii \varphi_{j} }\right) \right]^{w}.  \label{ZMM}
\end{eqnarray}
This matrix model, through its equivalent Toeplitz determinant formulation,
corresponds to the case of a pure Fisher-Hartwig (FH) singularity 
\cite{BS}. The transition between the exponentially decaying case above and
the long-range here could be itself studied through suitable double scaling limits
involving $\alpha $ (taking it $\alpha \rightarrow 0$) and $N$. This
transition involves going from Szeg\H{o} asymptotics to FH asymptotics, with
control over the transition through the solution of an integrable system 
\cite{P5}.

We focus here on the fact that \eref{ZMM} admits a remarkable exact
evaluation for arbitrary finite $N$, given by \cite{BS}%
\begin{equation}
Z(w)=e^{whN} \frac{G(N+1)G(2w+N+1)G(w+1)^{2}}{G(2w+1)G(N+w+1)^{2}},  \label{Zw2}
\end{equation}%
with $G$ denoting a Barnes $G$-function, which is a double Gamma function.
It is defined as $G(n)=\prod\nolimits_{m=0}^{n-2}m!$ for $n=1,2,...$ (for
the extension to the whole complex plane, one may use $G(z+1)=\Gamma \left(
z\right) G\left( z\right) $ with $G(1)=1$). As an open problem, it would be interesting to study if the known zeroes and poles 
of (\ref{Zw2}) could be interpreted as a signal of a DQPT in the sense of \cite{HPK,Heyl:2018abn}.

Because the Barnes $G$-function has Stokes lines on its imaginary axis, its asymptotic expansion involves Stokes phenomena and the appearance of exponentially small contributions in the asymptotic expansion \cite{Nemes,Berry}.


The exponentially improved asymptotic expansion of the Barnes $G$-function
reads \cite{Nemes}
\begin{eqnarray}
\fl \log G\left( {z+1}\right)  \sim \frac{1}{4}z^{2}+z\log \Gamma \left( {z+1}%
\right) -\left( {\frac{1}{2}z\left( {z+1}\right) +\frac{1}{{12}}}\right)
\log z-\log A \nonumber \\
+\sum\limits_{k=1}^{\infty }{S_{k}\left( \theta \right) e^{\pm
2\pi ikz}} +\sum\limits_{n=1}^{\infty }{\frac{{B_{2n+2}}}{{2n\left( {2n+1}%
\right) \left( {2n+2}\right) z^{2n}}}},  \label{asympt}
\end{eqnarray}%
where%
\begin{equation}
S_{k}\left( \theta \right) =%
\cases{
0 , & $ \left\vert \theta \right\vert <\frac{\pi }{2} $ \\ 
\mp \frac{1}{{4\pi ik^2 }} , & $ \theta =\pm \frac{\pi }{2} $ \\ 
\mp \frac{1}{{2\pi ik^2 }} , & $ \frac{\pi }{2}<\left\vert \theta \right\vert <\pi , $ \\ }
\label{eq25}
\end{equation}%
and $\theta =\arg z$. The upper or lower sign depends on whether $z$
is in the upper or lower half-plane. The term with $S_{k}\left( \theta
\right) $ are the Stokes multipliers. The remaining terms constitute
the usual asymptotic expansion of the Barnes $G$-function. The Stokes lines are
located at $\theta =\pm \frac{\pi }{2}$ in the complex plane.
Thus, this asymptotic expansion of the Barnes $G$-function on the
whole complex plane shows the existence of Stokes lines on the axis corresponding to real-time dynamics. Either expansions along the real-time line or crossings it, leads to the appearance of exponentially small contributions in the
expansion. The fact that these crossings and the appearance of the ensuing exponentially small extra terms in the asymptotic expansion are not actually a discontinuous behaviour is now understood and known as Berry smoothing
since in \cite{Berry}, zooming in the crossing regions, was shown that the process
is actually smooth, controlled by expressions characterized by error functions, instead of sharp transitions. 

Notice also that there is a Gamma function term in \eref{asympt},
therefore, we also have to consider the asymptotics of the Gamma function,
which also has Stokes lines in the same location, but with different Stokes
multipliers. More precisely: for the Gamma function the following asymptotic
expansion holds as $z\rightarrow \infty $%
\begin{equation}
\log \Gamma ^{\ast }(z)\sim \sum_{n=1}^{\infty }\frac{B_{2n}}{2n(2n-1)z^{2n-1}}-
\cases{
0 , & $ \left\vert \theta \right\vert <\frac{\pi }{2} $ \\ 
\frac{1}{2}\log (1-e^{\pm 2\pi iz}) , & $  \theta =\pm \frac{\pi }{2} $ \\ 
\log (1-e^{\pm 2\pi iz}) , & $ \frac{\pi }{2}<\left\vert \theta \right\vert <\pi . $ \\}
\end{equation}%
The expansion of the logarithm brings the asymptotics in the same form as
above%
\begin{equation}
\log \Gamma ^{\ast }(z)\sim \sum_{n=1}^{\infty }\frac{B_{2n}}{%
2n(2n-1)z^{2n-1}}+\sum\limits_{k=1}^{\infty }\widetilde{S}{_{k}\left( \theta
\right) e^{\pm 2\pi ikz},}  \label{Gammasymp}
\end{equation}%
in the sector $\left\vert \arg z\right\vert \leq \pi -\delta <\pi $ for any $0<\delta \leq \pi $ with\footnote{Note that, with regards to the location of Stokes lines, that the asymptotics of the Gamma function is with variable $z$ whereas of the Barnes $G$-function is $z+1$.}
\begin{equation}
\widetilde{S}_{k}\left( \theta \right) =
\cases{
0 , & $ \left\vert \theta \right\vert <\frac{\pi }{2} $ \\ 
\frac{1}{2k} , & $ \theta =\pm \frac{\pi }{2} $ \\ 
\frac{1}{k} , & $ \frac{\pi }{2}<\left\vert \theta \right\vert <\pi , $ \\ }
\end{equation}%
where the usual definition 
\begin{equation}
\Gamma ^{\ast }(z)=\frac{\Gamma (z)}{\sqrt{\pi }z^{z-1/2}e^{-z}}.
\end{equation}%
Notice that for real-time dynamics, while three Barnes $G$-function are on
the first quadrant of the complex plane and no exponentially small
contribution appears from those, the two terms $G(it+1)^{2}$ and $G(2it+1)$
are on the very Stokes line (for $t>0$, for $t<0$ it would be the
anti-Stokes line).


We can have actual crossings of the Stokes line in the case of complex time $w=\beta + \ii t \in \mathbb{C}$ and allowing $\beta <0$. On the other hand, for imaginary time asymptotics, when $w=\beta \in 
\mathbb{R}$ the phenomena does not occur, as all the asymptotics of \eref{Zw2} is
that of the Barnes $G$-function on the real axis. This then is an explicit
example where, under analytical continuation from imaginary-time to
real-time dynamics, we have additional subdominant exponentials, that would be missed in a direct Wick rotation from the asymptotics valid at the real axis (imaginary-time).

\subsection{A more general long and short range interaction. Phase transitions.}
\label{sec:GenIntP}

We consider a modification of the exponentially decaying interaction, corresponding to a
potential 
\begin{equation}
V(z)= h + \mathrm{Li}_{p+1}(e^{-\alpha} z)= h + \sum_{n=1}^{\infty }\frac{e^{- n \alpha} z^{n}}{n^{1+p}},
\end{equation}%
for $p\in \mathbb{Z}$ and $\alpha > 0$ when $p\leq 0$ and $\alpha \ge 0$
when $p>0$. At $p=0$ we recover \eref{Zw}. According to the general solution described in section \ref{app:saddleptsmethod}, this model has two phases, and the transition takes place at the critical value: 
\begin{equation}
\gamma _{c}=-\left( 2\mathrm{Li}_{p}(- e^{- \alpha} )\right) ^{-1}\geq 0.
\end{equation}%
The free energy in the weak coupling phase is: 
\begin{equation}
\mathcal{F}(\gamma \leq \gamma _{c})= \gamma h + \gamma ^{2}\mathrm{Li}_{2p+1}(e^{-2
\alpha}).
\label{FPolylogI}
\end{equation}%
At $p=0$ we recover formula \eref{FBaikIandII}.
On the other hand, for $p>0$ we can set $\alpha=0$, obtaining a 
long-range interaction which decays as $1/n^{(1+p)}$, corresponding to \eref{Jlmexp} with exponent $\lambda>1$. In this case $\mathcal{F}$ is proportional to the Riemann zeta function:
\begin{equation}
\mathcal{F}(\gamma \leq \gamma _{c})= \gamma h + \gamma ^{2}\zeta (2p+1).
\end{equation}

\begin{figure}[htb]
\centering
\includegraphics[width=8.6cm]{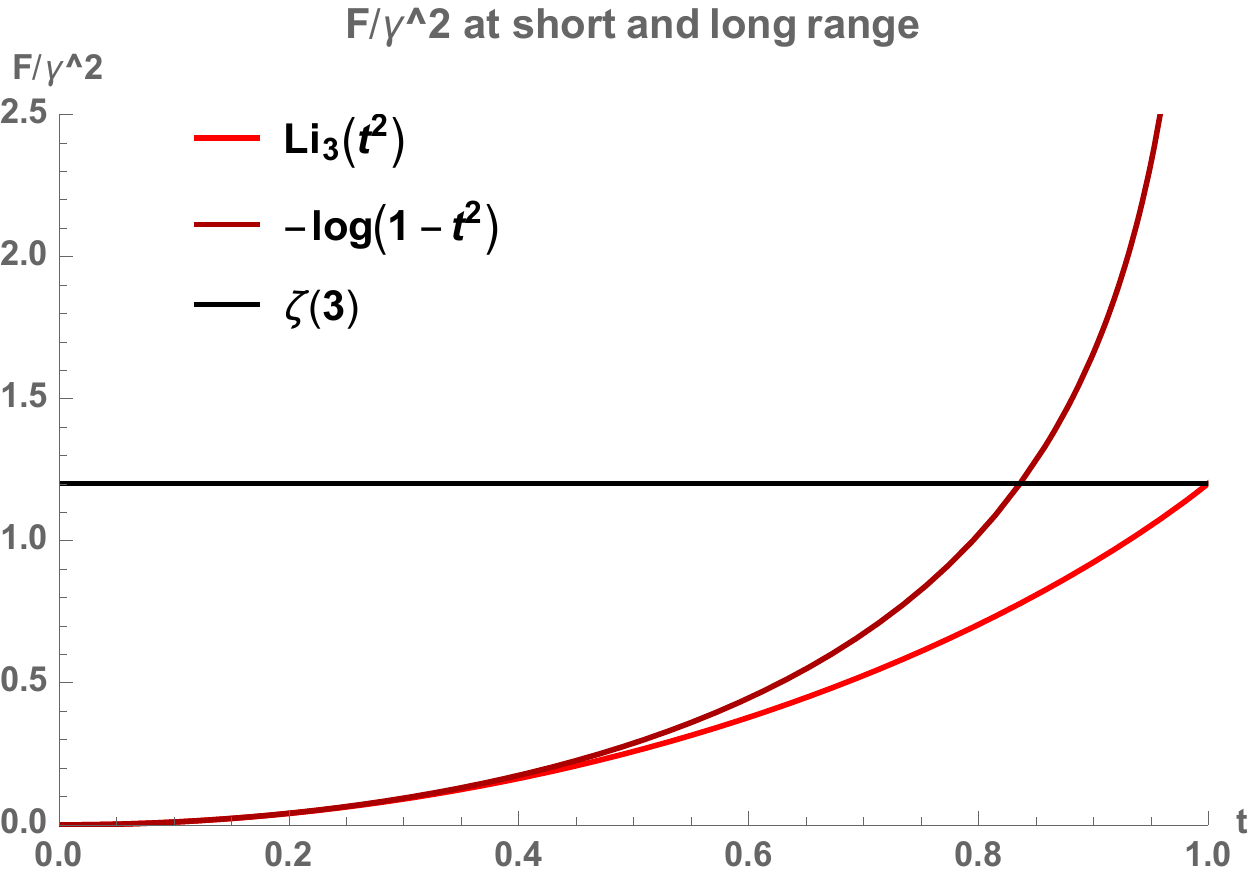}
\caption{Ratio between $\mathcal{F}$ and $\protect\gamma ^{2}$ as a function
of $t=e^{- \protect\alpha}$ for decays $e^{-n \protect\alpha}/n$ (red) $%
e^{-n \protect\alpha}/n^{2}$ (dark red) and $1/n^{2}$ (black). Plot at $h=0$. The case $p=0$
is singular at $t=1$.}
\end{figure}
\begin{figure}[htb]
\centering
\includegraphics[width=8.6cm]{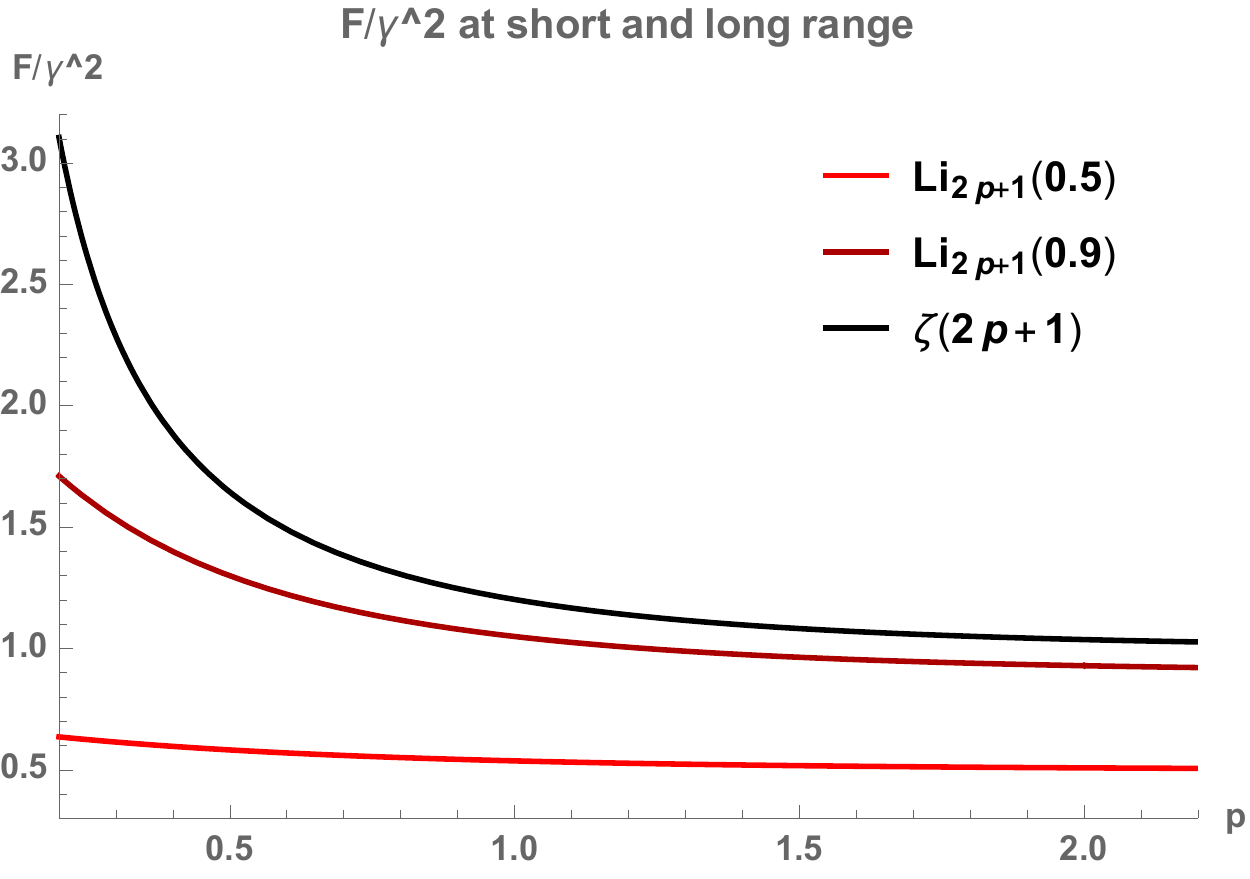}
\caption{Ratio between $\mathcal{F}$ and $\protect\gamma ^{2}$ as a function
of $p$ for decays $e^{- \alpha}=0.5$ (red) $e^{- \alpha}=0.9$ (dark red) and $e^{- \alpha}=1$ (black). Plot at $h=0$. For any value of $p$, $\mathcal{F}$ increases as $\alpha$ is decreased.}
\end{figure}
At $\gamma >\gamma _{c}$ the solution \eref{FPolylogI} ceases to be valid, and we ought to pursue a new one. For all $p \le 0$ the $\mathcal{F}$ is given by a simple modification of formula \eref{FBaikIandII}. For long-range interaction, $p>0$ and $\alpha =0$, we take advantage of a result in \cite{AmadoSundborg}. The formula for $\mathcal{F}$, however, is more involved and given by a double sum, presented in section \ref{app:saddleptsmethod}, Eq. \eref{dFdtPolylogII}.

\section{Finite chain}
\label{sec:finite}

To conclude our analysis, we now consider the case of a finite spin chain with $L+1$
sites, and focus on the extended model with exponentially decaying interaction discussed above. The
partition function \eref{Zw} is replaced by its discrete version: 
\begin{eqnarray}
\fl \mathcal{Z}_{\mathrm{d}}(\beta ) = \frac{1}{(L+1)^{2}}\sum_{0\leq
s_{1}<\dots <s_{N}\leq L}\ & \prod_{1\leq j<k\leq N} \vert e^{{\,\mathrm{i}\,}
\varphi _{s_{j}}}-e^{{\,\mathrm{i}\,}\varphi _{s_{k}}} \vert ^{2}  \nonumber \\
& \times \prod_{j=1}^{N}\left[ (1-e^{-\alpha }e^{\mathrm{i}\varphi
_{s_{j}}})(1-e^{-\alpha }e^{-\mathrm{i}\varphi _{s_{j}}})\right] ^{-\beta },
\label{ZBaikd}
\end{eqnarray}
with discrete angular variables already introduced in \eref{eq:discrangvars2} and defined as: 
\begin{equation}
\varphi _{s}=\frac{2\pi }{L+1}\left( s-\frac{L}{2}\right) .  \label{discrphi}
\end{equation}%
Mathematically this corresponds to discrete Toeplitz determinants \cite{discrete}. We take the large $N$ limit, with $\frac{\beta }{N}=\gamma
\geq 0$ and $\frac{L}{N}=r\geq 1$ fixed. The leading contribution to the
partition function in this limit is obtained from the same equation as for the continuous matrix model.
Nevertheless, the discrete matrix model at large $N$ is subject to an
additional constraint: the eigenvalues are distributed among the $L+1$ sites
on the circle, thus the distance between two of them is at least $\frac{2\pi }{L+1}$
 \cite{discrete,MinwallaWadia}. This follows from the ordering 
\begin{equation}
0\leq s_{1}<s_{2}<\dots <s_{N}\leq L,
\end{equation}%
together with the definition \eref{discrphi}, and hence we infer that the discrete eigenvalue density is further constrained by
\begin{equation}
 0 \le \rho_{\mathrm{d}} \le \frac{ r }{2 \pi} .
\end{equation}
Therefore the solution at weak coupling obtained in section \ref{app:saddleptsmethod}, Eq. \eref{rhoIBaik} holds only as long as both the upper and lower bound are satisfied. As a consequence, the discrete model presents two critical values of $\gamma $: 
\begin{equation}
\gamma_{c,0} =\frac{1+e^{- \alpha}}{2 e^{- \alpha}} , \quad \gamma_{c,r} =\frac{(1-e^{- \alpha})(r-1) }{2e^{- \alpha}}  .
\end{equation}%
The first is the same as in the continuous model, whilst the second arises from the discreteness. For a long chain 
\begin{equation}
	r \ge \frac{2}{1-e^{-\alpha}} ,
\end{equation}
it holds $\gamma_{c,0} \le \gamma_{c,r}$ and the first phase transition is induced by the same effect as for the infinite chain model. The free energy $\mathcal{F}$ when $\gamma \le \gamma_{c,r}$ is the same as in Eq. \eref{FBaikIandII}. As a check, we notice that in the limit $r\to\infty $ we consistently recover the infinite spin chain picture.

On the contrary, for a short chain, that is when $r < 2/(1-e^{-\alpha})$, the finite size effects appear earlier, the small coupling phase extends only up to $\gamma_{c,r}$ and the first phase transition is induced by discreteness effect.

In the context of the XX chain, a result from J. Baik and Z. Liu \cite{discrete} was used in \cite{DM} to show how, when the chain is longer than a threshold, $L> L_0 (\beta, N)$, the continuous matrix model (infinite chain) approximates the discrete one (finite chain) with exponentially small error. We now apply that method and show that the same holds for this other model.

\subsection{Comparison between finite and infinite chain}
\label{app:discreteMM}
We are interested in the ratio
\begin{equation}
\mathcal{R}_{L,N} (\beta) = \frac{ \mathcal{Z}_{\mathrm{d}}(\beta ) }{ \mathcal{Z} (\beta ) } ,
\end{equation}
which provides a measure of the accuracy of the infinite chain approximation for a finite chain of length $L+1$. In the definition, $\mathcal{Z}$ and $\mathcal{Z}_{\mathrm{d}}$ are the partiton functions of the continuous \eref{Zw} and discrete matrix model \eref{ZBaikd}, respectively.

In the unscaled limit, Baik and Liu \cite{discrete} proved that the ratio $\mathcal{R}_{L,N} (\beta)$ goes as:
\begin{equation}
\label{eq:BaikLiuestimate}
	\mathcal{R}_{L,N} (\beta) = 1 + \mathcal{O} (e^{- c N}) ,
\end{equation}
for some $c>0$, when the length of the chain is greater than a treshold, namely
\begin{equation}
L>L_0 (\beta, N) .
\end{equation}
This means that, when the spin chain is long enough, the continuous model (infinite chain) approximates the discrete one (finite chain) with exponentially small error.
Furthermore, in \cite{DM} it was showed that the same result holds in the scaling limit. There, following \cite{discrete} and a personal communication by J. Baik, the explicit expression of $L_0 (\beta, N)$ for the XX model in both phases was given. We apply the same calculations here: imposing $\rho_{\mathrm{d}} \le r/( 2 \pi)$ in both phases we obtain 
\begin{equation}
L_0 (\beta, N)  = \cases{ N  + \frac{ e^{-\alpha} }{1- e^{- \alpha}} \beta  ,& $  \gamma \le \gamma_{c,0} , $ \\ \frac{2}{1- e^{- \alpha}} \sqrt{e^{- \alpha} N \left( 2 \beta - N \right) } , & $  \ \gamma > \gamma_{c,0} . $ \\  }
\end{equation}

Let us remark that such estimates are obtained for the Hamiltonian with exponentially decaying interaction, $\alpha>0$, thanks to the result of \cite{discrete} on the asymptotic behaviour of discrete Toeplitz determinants with analytic symbol. Sending $\alpha \to 0$, the symbol develops a Fisher-Hartwig (FH) singularity. 
An interesting open problem would be to obtain the analogue of formula \eref{eq:BaikLiuestimate} when the symbol has FH singularities.

\ack
MT is indebted to David P\'{e}rez-Garc\'{\i}a for numerous stimulating discussions, over the years, and for collaboration on this subject. We thank also Bruno Mera and Riccardo Conti for useful discussions. The work of MT is supported by the Funda\c{c}\~{a}o para a Ci\^{e}ncia e a Tecnologia (FCT) through its program Investigador FCT IF2014, under contract IF/01767/2014. The work of LS is supported by the FCT through the doctoral scholarship SFRH/BD/129405/2017. The work is also supported by FCT Project PTDC/MAT-PUR/30234/2017.






\section*{References}

\begin{thebibliography}{99}
\bibitem{HPK} M. Heyl, A. Polkovnikov and S. Kehrein, \textquotedblleft
Dynamical Quantum Phase Transitions in the Transverse Field Ising
Model,\textquotedblright\ \href{https://journals.aps.org/prl/abstract/10.1103/PhysRevLett.110.135704}{Phys. Rev. Lett. \textbf{110}}, 135704 (2013),
\href{https://arxiv.org/abs/1206.2505}{[arXiv:1206.2505 [cond-mat.stat-mech]]}.

\bibitem{Heyl:2018abn} M.~Heyl, \textquotedblleft Dynamical quantum phase
transitions: a review,\textquotedblright\ \href{https://iopscience.iop.org/article/10.1088/1361-6633/aaaf9a/meta}{Rept.\ Prog.\ Phys.}\ \textbf{81},
no. 5, 054001 (2018) 
\href{https://arxiv.org/abs/1709.07461}{[arXiv:1709.07461 [cond-mat.stat-mech]]}. 


\bibitem{Piroli:2017mmz}  L.~Piroli, B.~Pozsgay and E.~Vernier,  ``From the
quantum transfer matrix to the quench action: the Loschmidt echo in XXZ
Heisenberg spin chains,''  \href{https://doi.org/10.1088/1742-5468/aa5d1e}{J.\ Stat.\ Mech.} \textbf{1702}, no. 2, 023106
(2017) 
\href{https://arxiv.org/abs/1611.06126}{[arXiv:1611.06126 [cond-mat.stat-mech]]}.

\bibitem{Stephan2017} J.M. St\'{e}phan, \textquotedblleft Return probability
after a quench from a domain wall initial state in the spin-1/2 XXZ
chain,\textquotedblright\ \href{https://iopscience.iop.org/article/10.1088/1742-5468/aa8c19}{J. Stat. Mech.} (2017) 103108, \href{https://arxiv.org/abs/1707.06625}{[arXiv:1707.06625
[cond-mat.stat-mech]]}.

\bibitem{PE} B. Pozsgay and V. Eisler, \textquotedblleft Real-time dynamics
in a strongly interacting bosonic hopping model: Global quenches and mapping
to the XX chain,\textquotedblright\ \href{https://iopscience.iop.org/article/10.1088/1742-5468/2016/05/053107}{J. Stat. Mech.} (2016) 053107,
\href{https://arxiv.org/abs/1602.03065}{[arXiv:1602.03065 [cond-mat.stat-mech]]}.


\bibitem{Piroli:2018amn} L.~Piroli, B.~Pozsgay and E.~Vernier,
\textquotedblleft Non-analytic behavior of the Loschmidt echo in XXZ spin
chains: Exact results,\textquotedblright\ \href{https://www.sciencedirect.com/science/article/pii/S055032131830172X}{Nucl.\ Phys.\ B} \textbf{933}, 454
(2018). 
\href{https://arxiv.org/abs/1803.04380}{[arXiv:1803.04380 [cond-mat.stat-mech]]}.




\bibitem{Piroli:2017sei} L.~Piroli, B.~Pozsgay and E.~Vernier,
\textquotedblleft What is an integrable quench?,\textquotedblright\ \href{https://www.sciencedirect.com/science/article/pii/S0550321317303413}{Nucl.\ Phys \ B} 
\textbf{925}, 362 (2017) 
\href{https://arxiv.org/abs/1709.04796}{[arXiv:1709.04796 [cond-mat.stat-mech]]}. 

\bibitem{BP} B.~Pozsgay, \textquotedblleft Dynamical free energy and the
Loschmidt-echo for a class of quantum quenches in the Heisenberg spin
chain,\textquotedblright\ \href{https://iopscience.iop.org/article/10.1088/1742-5468/2013/10/P10028}{J. Stat. Mech.} (2013) P10028, \href{https://arxiv.org/abs/1308.3087}{[arXiv:1308.3087
[cond-mat.stat-mech]]}.

\bibitem{Fzeroes} F. Andraschko and J. Sirker, \textquotedblleft Dynamical
quantum phase transitions and the Loschmidt echo: A transfer matrix
approach,\textquotedblright\ \href{https://journals.aps.org/prb/abstract/10.1103/PhysRevB.89.125120}{Phys. Rev. B \textbf{89}}, 125120 (2014),
\href{https://arxiv.org/abs/1312.4165}{[arXiv:1312.4165 [cond-mat.str-el]]}.

\bibitem{Echofermion} J. Viti, J.M. St\'{e}phan, J. Dubail and M. Haque,
\textquotedblleft Inhomogeneous quenches in a free fermionic chain: Exact
results,\textquotedblright\ \href{https://iopscience.iop.org/article/10.1209/0295-5075/115/40011}{EPL (Europhysics Letters)} \textbf{115} (4), 40011 (2016), \href{https://arxiv.org/abs/1507.08132}{[arXiv:1507.08132 [cond-mat.stat-mech]]}.

\bibitem{qreturn} P. L. Krapivsky, J. M. Luck and K. Mallick, \textquotedblleft
Quantum return probability of a system of N non-interacting lattice
fermions,\textquotedblright\ \href{https://iopscience.iop.org/article/10.1088/1742-5468/aaa79a}{J. Stat. Mech.} (2018) 023104, \href{https://arxiv.org/abs/1710.08178}{[arXiv:1710.08178
[cond-mat.mes-hall]]}.

\bibitem{Largedev} H. Moriya, R. Nagao and T. Sasamoto, \textquotedblleft Exact
large deviation function of spin current for the one dimensional XX spin
chain with domain wall initial condition,\textquotedblright\ \href{https://iopscience.iop.org/article/10.1088/1742-5468/ab1dd6}{J. Stat. Mech.} (2019) 063105,
\href{https://arxiv.org/abs/1901.07228}{[arXiv:1901.07228 [cond-mat.stat-mech]]}.

\bibitem{DM} D.~P\'{e}rez-Garc\'{\i}a and M.~Tierz, \textquotedblleft
Mapping between the Heisenberg XX Spin Chain and Low-Energy
QCD,\textquotedblright\ \href{https://journals.aps.org/prx/abstract/10.1103/PhysRevX.4.021050}{Phys.\ Rev.\ X \textbf{4}}, no. 2, 021050 (2014) \href{http://arxiv.org/pdf/1305.3877.pdf}%
{[arXiv:1305.3877 [cond-mat.str-el]]}.

\bibitem{Bog} N.M. Bogoliubov, ``XX0 Heisenberg chain and random walks,'' \href{https://doi.org/10.1007/s10958-006-0332-2}{J. Math. Sci.} \textbf{138}, 5636--5643 (2006).

\bibitem{Bog2} N.M. Bogoliubov,``Integrable models for vicious and friendly
walkers'', \href{https://doi.org/10.1007/s10958-007-0160-z}{J. Math. Sci.} \textbf{143}, 2729--2737 (2007).

\bibitem{BC} N.M. Bogoliubov and C. Malyshev, ``The correlation functions of
the $XXZ$ Heisenberg chain in the case of zero or infinite anisotropy, and
random walks of vicious walkers,'' \href{https://doi.org/10.1090/S1061-0022-2011-01146-X }{St. Petersburg Math. Jour.} \textbf{22}, 359--377 (2011), \href{https://arxiv.org/abs/0912.1138}{[arXiv:0912.1138 [cond-mat.stat-mech]]}.

\bibitem{B} N.M. Bogoliubov, A.G. Pronko and J. Timonen, \textquotedblleft
Scaling of many-particle correlations in a dissipative
sandpile,\textquotedblright\ \href{https://doi.org/10.1007/s10958-013-1256-2}{J. Math. Sci.} \textbf{190} : 3, 411–418 (2013), \href{https://arxiv.org/abs/1102.5639}{[arXiv:1102.5639 [cond-mat.stat-mech]]}.

\bibitem{Minor} D. Bump and P. Diaconis, \textquotedblleft Toeplitz
Minors,\textquotedblright\ \href{http://www.sciencedirect.com/science/article/pii/S0097316501932145}{ J. Comb. Theor. A97, 252 (2002)}.

\bibitem{RM} P. J. Forrester, \emph{Log-gases and random matrices},
Princeton University Press, Princeton, NJ, (2010).

\bibitem{PGT} D. P\'{e}rez-Garc\'{i}a and M. Tierz, \textquotedblleft Chern-Simons theory encoded on a spin-chain", \href{https://iopscience.iop.org/article/10.1088/1742-5468/2016/01/013103}{J. Stat. Mech.} (2016) 013103, \href{https://arxiv.org/abs/1403.6780}{[arXiv:1403.6780 [cond-mat.str-el]]}.

\bibitem{GW} D.~J.~Gross and E.~Witten, ``Possible Third Order Phase
Transition in the Large N Lattice Gauge Theory,'' \href{https://journals.aps.org/prd/abstract/10.1103/PhysRevD.21.446}{Phys.\ Rev.\ D} \textbf{21}%
, 446 (1980).

\bibitem{Wadia} S.~R.~Wadia, ``$N$ = Infinity Phase Transition in a Class of
Exactly Soluble Model Lattice Gauge Theories,'' Phys.\ Lett.\ \textbf{93B},
403 (1980).


\bibitem{BDJ} J. Baik, P. Deift and K. Johansson, \textquotedblleft On the
distribution of the length of the longest increasing subsequence of random
permutations ,\textquotedblright\ \href{https://www.ams.org/journals/jams/1999-12-04/S0894-0347-99-00307-0/}{J. Amer. Math. Soc.} \textbf{12}, 1119 (1999), \href{https://arxiv.org/abs/math/9810105}{[arXiv:math/9810105 [math.CO]]}.

\bibitem{BDS} J. Baik, P. Deift and T. Suidan, \emph{Combinatorics and Random Matrix Theory}, American Mathematical Society (2016).



\bibitem{MinwallaWadia} S.~Jain, S.~Minwalla, T.~Sharma, T.~Takimi,
S.~R.~Wadia and S.~Yokoyama, ``Phases of large $N$ vector Chern-Simons
theories on $S^2 \times S^1$,'' \href{https://doi.org/10.1007/JHEP09(2013)009}{JHEP} \textbf{1309}, 009 (2013), \href{https://arxiv.org/abs/1301.6169}{[arXiv:1301.6169 [hep-th]]}.

\bibitem{AmadoSundborg}  I.~Amado, B.~Sundborg, L.~Thorlacius and
N.~Wintergerst, \textquotedblleft Probing emergent geometry through phase
transitions in free vector and matrix models,\textquotedblright\ \href{https://doi.org/10.1007/JHEP02(2017)005}{JHEP} 
\textbf{1702}, 005 (2017) \href{https://arxiv.org/abs/1612.03009}{[arXiv:1612.03009 [hep-th]]}.



\bibitem{LSM} E. Lieb, T. Schultz, and D. Mattis, \textquotedblleft Two
soluble models of an antiferromagnetic chain,\textquotedblright\ \href{https://www.sciencedirect.com/science/article/abs/pii/0003491661901154}{Ann. Phys.}
\textbf{16}, 407 (1961).


\bibitem{Stanley} R.P. Stanley, \emph{Enumerative Combinatorics}, Vol 2.
Cambridge University Press (2001).


\bibitem{Glauber} R. J. Glauber, \textquotedblleft Time-Dependent Statistics
of the Ising Model,\textquotedblright\ \href{https://aip.scitation.org/doi/abs/10.1063/1.1703954}{J. Math. Phys.} \textbf{4}, 294 (1963).


\bibitem{Pol} H.-H. Tu, Y. Zhang and X.-L. Qi,  \textquotedblleft  Momentum polarization: an entanglement measure of topological spin and chiral central charge \textquotedblright\ \href{https://journals.aps.org/prb/abstract/10.1103/PhysRevB.88.195412}{Phys. Rev. B \textbf{88}}, 195412 (2013), 
\href{https://arxiv.org/abs/1212.6951}{[arXiv:1212.6951 [cond-mat.str-el]]}.


\bibitem{DGM} 
  D.~Garc\'{\i}a-Garc\'{\i}a and M.~Tierz,
  ``Matrix models for classical groups and Toeplitz$\pm $Hankel minors with applications to Chern-Simons theory and fermionic models,'' 
  \href{https://arxiv.org/abs/1901.08922}{[arXiv:1901.08922 [hep-th]]}.


\bibitem{Bose} S. Bose, \textquotedblleft Quantum communication through spin
chain dynamics: an introductory overview,\textquotedblright\  \href{https://www.tandfonline.com/doi/full/10.1080/00107510701342313}{Contemp. Phys.} 
\textbf{48}, 13 - 30, (2007), \href{https://arxiv.org/abs/0802.1224}{[arXiv:0802.1224 [cond-mat.other]]}.

\bibitem{Kay} A. Kay, \textquotedblleft A Review of Perfect State Transfer
and its Application as a Constructive Tool,\textquotedblright\ \href{https://www.worldscientific.com/doi/abs/10.1142/S0219749910006514}{Int. J. Quantum Inf.} 
\textbf{8} (2010), 641-676; \href{https://arxiv.org/abs/0903.4274}{[arXiv:0903.4274 [quant-ph]]}.


\bibitem{VZ} L. Vinet and A. Zhedanov, \textquotedblleft How to construct
spin chains with perfect state transfer,\textquotedblright\ \href{https://journals.aps.org/pra/abstract/10.1103/PhysRevA.85.012323}{Phys. Rev. A \textbf{85}},
012323 (2012) \href{https://arxiv.org/abs/1110.6474}{[arXiv:1110.6474 [quant-ph]]}.


\bibitem{Szegoth} G. Szeg\H{o}, ``On certain Hermitian forms associated with the Fourier
              series of a positive function,'' Comm. S\'{e}m. Math. Univ. Lund [Medd. Lunds Univ. Mat. Sem.],
              Tome Suppl\'{e}mentaire (1952), 228--238.
              
\bibitem{Franchini2016} 
  F.~Franchini,
  ``An introduction to integrable techniques for one-dimensional quantum systems,''
  \href{https://link.springer.com/book/10.1007/978-3-319-48487-7}{Lect.\ Notes Phys.}\  {\bf 940}, (2017)
  \href{https://arxiv.org/abs/1609.02100}{[arXiv:1609.02100 [cond-mat.stat-mech]]}.



\bibitem{JB} J. Baik, \textquotedblleft Random vicious walks and random matrices,\textquotedblright \href{https://doi.org/10.1002/1097-0312(200011)53:11<1385::AID-CPA3>3.0.CO;2-T}{Comm. Pure Appl. Math.} 
\textbf{53}:11 (2000), 1385-141, \href{https://arxiv.org/abs/math/0001022}{[arXiv:math/0001022 [math.PR]]}.



\bibitem{Godreche} C. Godr\`{e}che, \textquotedblleft Dynamics of the
directed Ising chain\textquotedblright, \ \href{https://iopscience.iop.org/article/10.1088/1742-5468/2011/04/P04005}{J. Stat. Mech.} (2011) P04005 \href{https://arxiv.org/abs/1102.0141}%
{[arXiv:1102.0141 [cond-mat.stat-mech]]}.

\bibitem{GodLuck} C. Godr\`{e}che, J. M. Luck, \textquotedblleft
Single-spin-flip dynamics of the Ising chain\textquotedblright, \ \href{https://iopscience.iop.org/article/10.1088/1742-5468/2015/05/P05033/meta}{J. Stat. Mech.} (2015) P05033 
\href{https://arxiv.org/abs/1503.01661}{[arXiv:1503.01661 [cond-mat.stat-mech]]}.

\bibitem{Bessel}  NIST Digital Library of Mathematical Functions, \href{https://dlmf.nist.gov/10}{Chapter 10, Bessel functions}.

\bibitem{PeriwalShevitz} 
  V.~Periwal and D.~Shevitz,
  ``Exactly Solvable Unitary Matrix Models: Multicritical Potentials and Correlations,''
  \href{https://www.sciencedirect.com/science/article/pii/0550321390906765}{Nucl.\ Phys.\ B {\bf 344}}, 731 (1990). 


\bibitem{LST} L. Santilli and M. Tierz, In preparation.

\bibitem{Copro} F. Colomo and A. G. Pronko. \textquotedblleft Third-order phase transition in random tilings,\textquotedblright \href{https://journals.aps.org/pre/abstract/10.1103/PhysRevE.88.042125}{Phys. Rev. E \textbf{88}} (2013), p. 042125, \href{https://arxiv.org/abs/1306.6207}{[arXiv:1306.6207 [math-ph]]}.


\bibitem{FDC} F. D. Cunden, P. Facchi, M. Ligab\'{o} and P. Vivo.
\textquotedblleft Universality of the third-order phase transition in the constrained Coulomb
gas". \href{https://iopscience.iop.org/article/10.1088/1742-5468/aa690c}{J. Stat. Mech.} 1705.5 (2017), p. 053303. \href{https://arxiv.org/abs/1702.05071}{[arXiv:1702.05071 [math-ph]]}.


\bibitem{Exp} R. Islam, C. Senko, W. C. Campbell, S. Korenblit, J. Smith, A.
Lee, E. E. Edwards, C. C. J.Wang, J. K. Freericks, and C. Monroe, \textquotedblleft Emergence
and Frustration of Magnetism with Variable-Range Interactions in a Quantum
Simulator," \href{http://science.sciencemag.org/content/340/6132/583}{Science 340, 583-587} (2013), \href{https://arxiv.org/abs/1210.0142}{[arXiv:1210.0142 [quant-ph]]}.

\bibitem{BS} A. Bottcher and B. Silbermann, \textquotedblleft Toeplitz matrices and
determinants with Fisher-Hartwig symbols". \href{https://www.sciencedirect.com/science/article/pii/0022123685900850}{J. Funct. Anal.} \textbf{63} (1985), 178--214.


\bibitem{P5} T. Claeys, A. Its and I. Krasovsky, \textquotedblleft Emergence of a singularity
for Toeplitz determinants and Painleve V," \href{https://projecteuclid.org/euclid.dmj/1319721312}{Duke Math. J.} \textbf{160} (2011),
207-262, \href{https://arxiv.org/abs/1004.3696}{[arXiv:1004.3696 [math-ph]]}.


\bibitem{Nemes} G. Nemes, \textquotedblleft Error bounds and exponential
improvement for the asymptotic expansion of the Barnes
G-function,\textquotedblright\ \href{https://royalsocietypublishing.org/doi/10.1098/rspa.2014.0534}{Proc. R. Soc.} A \textbf{470}, (2014) \href{https://arxiv.org/abs/1406.2535}%
{[arXiv:1406.2535 [math.CA]]}.

\bibitem{Berry} M. V. Berry, ``Uniform asymptotic smoothing of Stokes's
discontinuities,'' \href{https://royalsocietypublishing.org/doi/abs/10.1098/rspa.1989.0018}{Proc. R. Soc.} A \textbf{422}, 7-21 (1989).


\bibitem{discrete} J. Baik and Z. Liu, \textquotedblleft Discrete Toeplitz/Hankel Determinants and the Width of Nonintersecting Processes \textquotedblright\ 
\href{ https://doi.org/10.1093/imrn/rnt143}{Int. Math. Res. Not.} \textbf{20} (2014), 5737--5768,  \href{https://arxiv.org/abs/1212.4467}{[arXiv:1212.4467 [math.PR]]}.


\end{thebibliography}
\end{document}